\documentclass[12pt,a4wide]{article}
\usepackage{axodraw,amssymb,amsmath,color}
\usepackage[bookmarks]{hyperref}
\usepackage[mathcal]{eucal}

\def\mA{\ensuremath{\mathcal{A}}}

\def\mG{\ensuremath{\mathcal{G}}}

\def\mL{\ensuremath{\mathcal{L}}}

\newcommand{\cL}{\ensuremath{{\cal{L}}}}
\newcommand{\cA}{\ensuremath{{\cal{A}}}}
\newcommand{\cB}{\ensuremath{{\cal{B}}}}

\newcommand{\cO}{\ensuremath{{\cal{O}}}}

\def\nn{\nonumber \\}
\def\fs{\; \; .}
\def\co{\; \; ,}

\def\eps{\ensuremath{\varepsilon}}

\def\fs{\; \; .}
\def\co{\; \; ,}

\def\cpt{\mbox{CHPT\,}}

\def\bdm{\begin{displaymath}}
\def\edm{\end{displaymath}}
\def\be{\begin{equation}}
\def\ee{\end{equation}}

\def\bea{\begin{eqnarray}}
\def\eea{\end{eqnarray}}
\def\bes{\begin{eqnarray*}}
\def\ees{\end{eqnarray*}}

\def\ba{\begin{array}}
\def\ea{\end{array}}
\def\bal{\begin{align}}
\def\eal{\end{align}}
\def\bspl{\begin{split}}
\def\espl{\end{split}}
\def\bml{\begin{multline}}
\def\eml{\end{multline}}
\def\no{\nonumber}

\def\bdm{\begin{displaymath}}
\def\edm{\end{displaymath}}

\renewcommand{\theequation}{\arabic{equation}}

\def\ho{$ \hbar $-order }
\def\lo{$ l $-order }
\def\vo{$ v $-order }

\renewcommand{\theequation}{\arabic{section}.\arabic{equation}}

\begin{document}
\bibliographystyle{aip}

 \title{Renormalization group equations \\ for effective field theories}
\author{ 
 M.~B\"uchler$^{a,b}$ and G.~Colangelo,$^b$ \vspace{1cm}\\
{\small
${}^a$ Institut f{\"u}r Theoretische Physik, Universit{\"a}t Z\"urich}\\
{\small Winterthurerstr. 190, 8057 Z\"urich,  Switzerland}  \\
{\small
${}^b$ Institut f{\"u}r Theoretische Physik, Universit{\"a}t Bern}
\\
{\small Sidlerstr. 5, 3012 Bern,  Switzerland } }
\maketitle
\begin{abstract}
We derive the renormalization group equations for a generic
nonrenormalizable theory. We show that the equations allow one to derive
the structure of the leading divergences at any loop order in terms of
one-loop diagrams only. In chiral perturbation theory, e.g., this means
that one can obtain the series of leading chiral logs by calculating only
one loop diagrams. We discuss also the renormalization group equations for
the subleading divergences, and the crucial role of counterterms that
vanish at the equations of motion.  Finally, we show that the
renormalization group equations obtained here apply equally well also to
renormalizable theories.
\end{abstract}
\thispagestyle{empty}
\setcounter{page}{0}

\newpage
\tableofcontents
\newpage
\section{Introduction}

Quantum field theories (QFT) which are used in phenomenology are tested up
to a limited level of precision and in limited ranges of energies.  In their
formulation and application one does not need to worry about if and how the
theory has to be modified once certain boundaries in energy (or precision)
are crossed: in such cases one usually speaks about effective field
theories. The property of renormalizability of such quantum field theories
is conceptually not particularly relevant\footnote{This is the point of
view of many modern textbooks on quantum field theories. See, e.g.,
Ref.~\cite{WeinbergQFT,PeskinQFT}.} -- at most one can work out predictions
to an interesting level of precision using only the renormalizable part of
the interaction Lagrangian. The latter is the case of the Standard Model,
where the level of precision which has been reached without need for new
nonrenormalizable interactions has become surprisingly high.

In renormalizable quantum field theories one of the most useful tools
is that of the renormalization group equations (RGE). After the
renormalization procedure, the coupling constants which define the theory
acquire a dependence on an arbitrary energy scale. It is convenient to
identify the latter with the typical energy scale of the process under
consideration -- the strength of the interaction then varies with the
energy at which this occurs. The RGE dictate how the coupling constants
depend on the scale, and are one of the most important intrinsic properties
of a quantum field theory. As is well known, the discovery of the property
of asymptotic freedom for nonabelian gauge theories was a major
breakthrough, and showed that such theories could be candidates for
describing the observed behaviour of hadrons in deep inelastic scattering,
which then led to the formulation of QCD.

The use of the RGE in nonrenormalizable QFT has not received the same
attention, and has not yet been studied thoroughly. On one hand this may be
due to the different use of nonrenormalizable QFT, where one usually does
not have the problem of evolving coupling constants over order of
magnitudes in energy scales. On the other hand, the very structure of the
RGE in the case of nonrenormalizable QFT is a lot more complicated than for
renormalizable ones. One of the very first investigations of this issue was
made by Weinberg \cite{Weinberg:1979kz}, in his seminal paper on the
effective Lagrangians. There he shows how one can follow
the same reasoning that leads to the RGE for renormalizable theories to
obtain informations about the structure of the two-loop divergences in
chiral perturbation theory (CHPT). He does not attempt, however, to push
the analysis to higher orders. This is the aim of the present paper.

What is the physical information one would like to obtain from such an
analysis? To illustrate the answer let us consider, for example, the
expansion of the pion mass in quark masses \cite{Colangelo:1995jm}:
\bea
M_\pi^2 &=& M^2 \left[ 1 + {M^2 \over (4 \pi F)^2} \left(- {1 \over 2} 
\log {M^2 \over \mu^2} + \ell_3^r(\mu) \right) \right. \nn 
&+& \left.  {M^4 \over (4 \pi F)^4} \left({17 \over 8} \log^2 {M^2
    \over \mu^2} + \ldots \right) +O(M^6) \right] \co
\eea
where $M^2=2 \langle \bar q q \rangle \hat m/F^2 $ is the
Gell-Mann--Oakes--Renner term, and $F$ the pion decay constant in the
chiral limit. We have stopped the expansion at the next-to-next-to-leading
order, and at this order have written down explicitly only the double
chiral logarithm. Indeed, what we want to show here, is that the
coefficient of the single (double) chiral logarithm at order $M^4$ ($M^6$)
is a pure number, and does not involve any of the new coupling constants
that show up at each order in the chiral expansion. It follows from simple
power counting that this remains true to all orders: the coefficient of the
leading chiral log at any order in the chiral expansion is a pure
number. The analysis of Weinberg \cite{Weinberg:1979kz} concerned precisely
the coefficient of the double chiral log: he showed that, although in
principle that coefficient is the sum of contributions of several different
loop diagrams, its value is constrained by the RGE, and can in fact be
obtained from one-loop diagrams only \cite{Colangelo:1995jm}. Since the
leading chiral log is potentially the dominating correction at each chiral
order (this statement is of course $\mu$-dependent, and typically valid
for values of $\mu$ around 1 GeV), the RGE do provide information of
phenomenological interest. If one were able to sum the whole series of the
leading chiral logs with the help of the RGE, this would certainly be a
very useful and exciting result.

Analogous statements are true for any observables, and indeed one can
obtain the complete expression of the double chiral logs from the RGE in
the generating functional \cite{Bijnens:1998yu}, in the formulation of CHPT
with external fields which is due to Gasser and Leutwyler
\cite{Gasser:1984yg}. The validity of the RGE at the two-loop level has
then been explicitly verified in \cite{Bijnens:1999hw} in the full
two-loop calculations of the divergence structure of CHPT. The extension
of these RGE arguments to higher orders has however not yet been made in
the framework of CHPT.

The RGE in a nonrenormalizable QFT have been studied for the case of the
nonlinear $\sigma$-model in two dimensions
\cite{Friedan,Alvarez-Gaume}. This low-dimensional QFT is particularly
interesting because, on one hand, if one does not specify the metric of the
manifold on which the fields live, it is nonrenormalizable. But on the
other hand, the structure of the possible counterterms is severely
constrained: only two space-time derivatives of the fields can appear, such
that the counterterms can always be absorbed in a redefinition of the
metric. As was shown by Friedan \cite{Friedan} one can write down the RGE
for the metric, which do imply interesting constraints on the form of the
leading divergences at higher orders in the loop expansion.  This was
further analyzed and clarified by Alvarez-Gaum\'e, Freedman and Mukhi
\cite{Alvarez-Gaume}, who showed that on the basis of the RGE one can
derive the leading two-loop divergences from purely geometrical
considerations (the Palatini identity). They then verified that the actual
two-loop calculation gave results in agreement with the RGE.  In that
paper the RGE for the leading divergences were derived to all orders
(although in a rather implicit form).  A few years later, Kazakov
\cite{Kazakov:1988jp} extended these ideas to arbitrary QFT in four
space-time dimensions. He relied, however, on a specific assumption on the
scaling of the Lagrangian with the renormalization scale $\mu$ (the RGE
were derived in dimensional regularization), and the RGE were also given in
a very implicit form, completely analogous to those for the metric in the
2-dimensional nonlinear $\sigma$-model. As we will show later, however,
this scaling cannot hold in CHPT, and in our analysis we have had to adopt
a different starting point.

The structure of the paper is as follows: in Sect.~2 we set our notation
and derive the implicit form of the RGE. In Sect.~3 we analyze the RGE for
the leading divergences, first order by order in the loop expansion, and
then give the explicit all-order formula and discuss its meaning.  In
Sect.~4 we consider the RGE for the subleading divergences, and discuss
them for the first few orders in the loop expansion. In Sect.~5 we discuss
the role of the counterterms that vanish at the solution of the equations
of motion, in connection with the role of one-particle-reducible graphs.
In Sect.~6 we show that the RGE we have derived apply equally well to the
case of a renormalizable QFT, and explicitly discuss the case of the $O(N)$
invariant $\phi^4$ theory. In that framework we can also illustrate the
role of the counterterms that vanish at the solution of the equations of
motion. In Sect.~7 we add a $\phi^6$ interaction to the $\phi^4$ theory and
analyze how the RGE are affected.  Finally we summarize our results in
Sect. \ref{sec:conc}.  In the appendices we discuss the more technical
points, and in particular the derivation of the RGE to all orders.

\setcounter{equation}{0}
\section{Renormalization group equations}
\label{sec:RGE}

\subsection{Notation}
\label{sec:notation}
A quantum field theory is defined by specifying its classical action $S_0$
and a series of quantum corrections $S_i$
\be
S[\phi,J] = \sum_{n=0}^{\infty} \hbar^{n} S_n[\phi,J]
\fs
\ee
Each term in the series is a function of a number of fields, collectively
denoted with the symbol $\phi$, and of external sources $J$ coupled to
operators $O$ (which can be either fields $\phi$ or functions thereof). By
evaluating the path integral 
\be
e^{iZ[J]/\hbar}:= {1 \over {\cal N}} \int \prod \left[d \phi_i\right]
e^{iS[\phi,J]/\hbar} \co
\ee
one obtains the generating functional $Z[J]$ of all the Green
functions of the operators $O_J$ as a power series in $\hbar$,
\be
Z[J] = \sum_{n=0}^{\infty} \hbar^{n} Z_n[J] \fs
\ee
In the evaluation of
the path integral divergences are generated: these need to be renormalized
in order to have physically meaningful results. This can be done in the
following way. The action $S$ is the integral over spacetime of the bare
Lagrangian which also admits an expansion in a power series in $\hbar$
\be
S[\phi,J]= \int dx  \mL^{\mbox{\tiny{bare}}}(\phi,J) \co \qquad
\mL^{\mbox{\tiny{bare}}} = \sum_{n=0}^{\infty} \hbar^{n} \mL^{(n)
    \mbox{\tiny{bare}}}  \fs
\ee
We regularize the theory by working in $d$ spacetime dimension, and
split the bare Lagrangians into a renormalized and a divergent part: 
\be
 \mL^{\tiny{(n) \mbox{bare}}}:= \mu^{-\eps n} (\mL^{\tiny{(n)}}+
 \mL^{\tiny{(n) \mbox{div}}})  \qquad n \geq 0
\label{eq:Lbare}
\co
\ee
where $ \mL^{\tiny{(n)}} $ is the renormalized Lagrangian, $\eps:=4-d$, and  $\mL^{\tiny{(n) \mbox{div}}} \quad n \geq 1$ diverges in the
limit $\eps \to 0$. All the divergences generated in the calculation of the
path integrals are local (see, e.g. \cite{WeinbergQFT}), and can be
reabsorbed by properly defining $\mL^{\tiny{(n) \mbox{div}}}$.

The scale $\mu$ introduced in Eq.~(\ref{eq:Lbare}) serves the purpose of
having a renormalized Lagrangian of dimension $d$ for all
$\hbar$-orders. The reason why this choice is the correct one in CHPT is
explained in Appendix~\ref{ap:diman}. In case of a renormalizable
Lagrangian other choices would be more appropriate -- on the other hand,
the physical content of the RGE does not depend on this, as we will see in
Sect.~\ref{sec:ren}. From now on we set $\hbar=1$.  In the framework of
\cpt the renormalized Lagrangian $ \mL _{n} $ corresponds to the Lagrangian
of chiral order $ 2n+2 $.

The divergent part of the bare Lagrangian of \ho $n$ can be written as a
sum of poles in $\eps$
\be
 \mL^{\tiny{(n)\mbox{div}}} :=  \sum_{k=1}^{n} A^{(n)}_{k}\eps^{-k} =
 \sum_{k=1}^{n}\sum_{l=k}^n \cA^{(n)}_{lk}\eps^{-k} 
 \co
\ee
where after the second equality sign we have expanded the divergences in
terms generated by diagrams with $l$ loops -- obviously a term diverging
like $\eps^{-k}$ can only be generated by diagrams with at least $k$ loops.
The part of \ho $n$ of the bare Lagrangian therefore reads
\be
 \mL^{(n)\mbox{\tiny{bare}}} = \mu^{-\eps n} 
 \left[ \mL^{(n)} + \sum_{k=1}^{n} A^{(n)}_k \eps^{-k} \right] \fs
\label{eq:Lnbare}
\ee
The calculation of the divergent coefficients $A^{(n)}_k$ can be performed
in various different ways, which we need not specify here. The use of the
background field method and the heat-kernel techniques are particularly
convenient in cases where a local symmetry is present, like for gauge
theories or CHPT. Concrete examples of calculations of $A^{(n)}_k$ up to
$n=2$ for gauge theories and CHPT can be found in \cite{Jack:1982hf} and
\cite{Bijnens:1999hw}, respectively.

At each \ho both the Lagrangian $\cL^{(n)}$ and the pole coefficients
$A^{(n)}_k$ can be expanded in a minimal basis of operators
$\cO^{(n)}_i$, $i=1,\ldots,M_n$:
\bea
\mL^{(n)}&=&\sum_{i=1}^{M_n} c^{(n)}_i \cO^{(n)}_i = \vec{c}^{(n)} \cdot
\vec{\cO}^{(n)}  \nn
A^{(n)}_k&=& \sum_{i=1}^{M_n} a^{(n)}_{k\,i} \cO^{(n)}_i = \vec{a}^{(n)}_k
\cdot \vec{\cO}^{(n)} = \sum_{l=k}^n \vec{a}^{(n)}_{lk} \cdot
\vec{\cO}^{(n)} \fs  
\label{eq:Ank}
\eea
For a renormalizable theory $M_n$ is a constant independent of $n$, and the
minimal basis of operators is the same for every $n$, whereas for a
nonrenormalizable one $M_n$ is a growing function of $n$. In the present
formalism this is the only difference between a renormalizable and a
nonrenormalizable theory.

\subsection{Renormalization group equations}
The RGE follow from the requirement that the bare Lagrangians\footnote{From
  now on we suppress the superscript $r$ on the renormalized Lagrangian.} 
(\ref{eq:Lnbare}) do not depend on the scale $\mu$: 
\bea
 0 & = & \mu \frac{d}{d \mu} \mL^{(n) \mbox{\tiny{bare}}}
\\
 & =& \mu^{-\eps n} \left\{ -\eps n \left[ \mL^{(n)} +
\sum_{k=1}^{n} \eps^{-k} A^{(n)}_k \right] + \mu \frac{d}{d \mu}
\mL^{(n)} + \sum_{k=1}^{n} \eps^{-k}  \mu \frac{d}{d \mu} A^{(n)}_k
\right\} \nonumber \fs
\label{eq:poe1}
\eea
The $ \mu $ dependence of the $ \mL^{(n)} $ will be described by its
$\beta$-function, denoted by the symbol $\cB$: 
\be
 \mu \frac{d}{d \mu} \mL^{(n)} = \cB^{(n)} + \eps n \mL^{(n)}  \co
\label{eq:beta1}
\ee
where the $ \eps $-dependence has been explicitly subtracted.
The $\beta$-function of a Lagrangian is also expandable in the set of 
operators:
\be
 \cB^{(n)} = \sum_{i=1}^{M_n} \beta^{(n)}_i \cO^{(n)}_i = \vec{\beta}^{(n)}
 \cdot \vec{\cO}^{(n)} \co
\ee
and is defined to be evaluated at $d=4$. If we decompose
Eq.~(\ref{eq:beta1}) into a basis of operators we obtain, for each of the
coupling constants $c^{(n)}_i$
\be
 \mu \frac{d}{d \mu} c^{(n)}_i = \beta^{(n)}_i + \eps n c^{(n)}_i \fs
\label{eq:beta2}
\ee

The $\mu$ dependence of the divergent parts $A^{(n)}_k$
can only arise through their explicit polynomial dependence on the coupling
constants $c^{(n)}_i$. We can therefore rewrite the $\mu$ derivative as
\be
\label{eq:md2}
\mu\frac{d}{d \mu} = \sum_n \mu {d \vec{c}^{(n)} \over d \mu} 
\cdot \vec{\partial}^{(n)} =\sum_{n} \left[
  \vec{\beta}^{(n)} + \eps n \vec{c}^{(n)} \right] \cdot
\vec{\partial}^{(n)}= \nabla+\eps N_v 
\ee
where $\vec{\partial}^{(n)}:= \partial / \partial \vec{c}^{(n)}$, and we
have introduced the following definitions: 
\bea
\nabla&:=& \sum_n \nabla^{(n)} \co \qquad \nabla^{(n)}:= \vec{\beta}^{(n)}
\cdot \vec{\partial}^{(n)} \nn 
N_v &:=& \sum_n n D^{(n)}\co \qquad D^{(n)} := \vec{c}^{(n)} \cdot
\vec{\partial}^{(n)} 
\label{eq:nv}
\eea

The operator $N_v$ and its eigenvalues have a clear meaning which we are
now going to illustrate. We saw above that every Lagrangian $\mL^{(j)} $
comes with a factor $\hbar^{j}$. Consider a generic object on which the
operator $N_v$ will act, e.g. an $A^{(n)}_k$, and more in particular the
contribution of a specific loop graph to it, denoted by $\mG$. When
acting on a diagram, the operator $D^{(j)}$ will yield the number $n_j$ of
vertices coming from the Lagrangian $\mL^{(j)}$ which are present in that
diagram:
\be
D^{(j)} \mG := n_j \mG \fs
\ee
For $ N_{v} $ we therefore get: 
\be
 N_{v} \mG = \sum j n_{j} \mG =: n_{v} \mG \co
\ee
where $n_{v}$ (called \vo) is the contribution to the \ho of a diagram
$\mG$ which is coming only from the vertices. The total \ho of the diagram
must be larger or equal to $n_{v}$, and the difference between $ n $ and $
n_{v} $ is generated dynamically by the $n_{l}$ loops present in that
diagram: $n=n_{v}+n_{l}$ (for $n_l$ we will use the term \lo). 
As already mentioned above, for a renormalizable theory
we would have $ n_{v} \equiv 0 $ and therefore $n = n_{l}$. 

We can now write Eq. (\ref{eq:poe1}) in a very compact form:  
\bea
\cB^{(n)} & = & N_{l} A^{(n)}_1 \co \label{eq:sdkaz1} \\
 N_{l} A^{(n)}_{k+1} & = & \nabla A^{(n)}_{k} \qquad \qquad \qquad k =
 1,...,n-1 
\label{eq:sdkaz2} \co
\eea
where $N_l$ is the operator that yields the $l$-order of the object it
acts on, and is defined as 
\be
N_{l} := n - N_{v} \fs
\ee
We observe that terms of different $l$-order in
Eqs.~(\ref{eq:sdkaz1},\ref{eq:sdkaz2}) cannot mix with each other.  One way
to prove this statement is the following: all the objects appearing in
Eqs.~(\ref{eq:sdkaz1},\ref{eq:sdkaz2}) are polynomials in the coupling
constants $c^{(n)}_i$. Since these identities hold no matter what the value
of these constants is, they must hold for the coefficients of each monomial
in the coupling constants. One can now group together all monomials with
the same $n_v$, which also have the same $l$-order $l=n-n_v$. This follows
from the fact that the \ho $n$ is constant for all terms in
Eqs.~(\ref{eq:sdkaz1},\ref{eq:sdkaz2}).

The RGE can therefore be decomposed into sets of equations with fixed
$l$-order
\bea
 \cB_{l}^{(n)} & = & l \cA_{l1}^{(n)} \qquad \qquad \quad \qquad l=1, \ldots n
 \co  \label{eq:k1l}\\ 
 l \cA_{lk}^{(n)} & = & \sum_{l'=1}^{l-k+1} \nabla_{l'} \cA_{l-l' \,k-1}^{(n)}
 \label{eq:k2l} \qquad l=k, \ldots n, \; \; k = 2,\ldots,n\co  
\eea
where the additional index $l$ stands for the loop order, and where $
\nabla_{l} $ is defined as: 
\be
 \nabla_{l} := \sum_{n=l}^{\infty}
 \vec{\beta}^{(n)}_{l} \cdot \vec{\partial}^{(n)}  \fs
\ee
The boundaries in the sum follow from the trivial observation that
$A^{(n)}_k$ has $l$-order $\ge k$ -- the \ho is of course equal to $n$. 

In order to further manipulate the RGE it is useful to establish the
following simple rules:
\begin{enumerate}
\item
$\cB^{(n)}$ and $ \mA^{(n)} $ can carry any \lo and \vo which add up to $n$. 
The $c_{i}^{(n)}$ have by definition: $ N_{l}c_{i}^{(n)} = 0, \; \;
N_{v}c_{i}^{(n)} = n$. A  derivative $\partial_{i}^{(n)}$ reduces the \vo
of the object it acts on by $n$.
\item
With the action of $ \nabla_{l}$ we differentiate with
$\vec{\partial}^{(n)} $ and multiply the result with the corresponding 
$\vec{\beta}_{l}^{(n)} $. The net change in the \ho $n$ is therefore
zero: $\nabla_{l} $ increases (decreases) the \lo (\vo)
of the object it acts on by $l$: 
\bea
 N_{l}(\nabla_{l_{1}} \mA_{l'k}^{(n)}) &=& (l_{1} +
 l')\nabla_{l_{1}}\mA_{l'k}^{(n)}   \nn \co
\; \;
 N_{v}(\nabla_{l_{1}} \mA_{l'k}^{(n)}) &=& (n -
 l_{1}-l')\nabla_{l_{1}}\mA_{l_{2}k}^{(n)}  \no \fs
\eea
\item
If we have not enough $ c_{i}^{(k)} $ inside of $ \mA_{l'k}^{(n)} $ on which 
we can act with the derivatives $ \vec{\partial}^{(n)} $, the $ \nabla_{l}
\mA_{l'k}^{(n)} $ will evaluate to zero:   
\be
 \nabla_{l} \mA_{l' k}^{(n)} = 0 \quad ; \; \forall l + l' > n \fs
 \no
\ee
\end{enumerate}
The above statements are valid also for products of $\nabla_{l}$ if we
substitute 
\be
\nabla_{l} \to \nabla_{l_{1}}\nabla_{l_{2}}...\nabla_{l_{k}} \co  \qquad
 l  \to  l_{1} + l_{2} + ...+ l_{k} \fs
\ee

\setcounter{equation}{0}
\section{RGE for the highest poles}
In this section we analyze in detail the RGE (\ref{eq:k2l}) for the highest
poles (HPRGE) $k=n$ and write them in a more compact form. Before getting to
the final result for generic $n$, we find it instructive to examine a first
few explicit cases starting from $n=1$. 

\subsection{RGE for $n=1, \; 2$ and $3$}
At lowest \ho the RGE are practically trivial:
\be
\cB^{(1)}_{1}= \cA^{(1)}_{11} \co
\label{eq:rge_l1}
\ee
and only state that the scale dependence of the couplings in the $\cL_1$
Lagrangian is determined by the one-loop divergences 
\cite{Weinberg:1979kz,Gasser:1984yg,Gasser:1985gg}. 

At the two loop level the equations become more interesting, as has been
already observed by Weinberg \cite{Weinberg:1979kz} and others
\cite{Colangelo:1995jm,Bijnens:1998yu,Bijnens:1999hw}:
\bea
\cB_{1}^{(2)}&=&   \cA^{(2)}_{11}  \co \nn
\cB_{2}^{(2)}&=& 2 \cA^{(2)}_{21}  \co \nn
2 \cA_{22}^{(2)}&=& \nabla_1 \cA^{(2)}_{11} = \nabla_1 \cB^{(2)}_{1} \fs
\label{eq:rge_l2}
\eea
In this case the RGE show that the scale dependence of the couplings in
$\cL_2$ is fully contained in the single pole in $\eps$,
$\cA^{(2)}_{21}$. Acting with $ N_{l} $ on $A^{(2)}_{1}$ yields
the two terms $ \cA^{(2)}_{11} + 2 \cA^{(2)}_{21} $, where the first will be
linear in the couplings coming from $ \mL^{(1)} $, and the latter will depend
only on $ \mL^{(0)} $.

The first of these equations is identical to (4.5) in \cite{Bijnens:1998yu},
whereas the second to (4.6) or (2.44) also in \cite{Bijnens:1998yu}.
As has already been observed
\cite{Weinberg:1979kz,Colangelo:1995jm,Bijnens:1998yu}  
the second of these equations allows one to calculate the double chiral
logs only with one-loop calculations.
In passing we note that $\nabla^{(1)}_1$ can also be written as
\be
\nabla^{(1)}_1 =  \vec{a}^{(1)}_{11} \cdot \vec{\partial}^{(1)}
\co
\label{eq:nabla1}
\ee
as follows from (\ref{eq:rge_l1}). For later convenience we introduce the
symbol: 
\be
d_n:= \vec{a}^{(n)}_{nn} \cdot \vec{\partial}^{(n)}
\label{eq:def_d}
\ee
With this notation (\ref{eq:nabla1}) can be re-expressed as $\nabla_1=d_1$.
We stress that all the $d_n$'s commute:
\be
[d_n,d_m]=0 \fs
\ee

\begin{figure}
\begin{eqnarray*}
\nonumber
 2 \qquad
 \begin{picture}(40,40)(0,-2)
 \CBox(-12,-12)(12,12){Black}{Yellow}
 \Text(0,0)[]{$\cA^{(2)}_{22}$}
\end{picture}
& = &
 \begin{picture}(40,40)(-40,-2)
 \CArc(25,0)(25,0,360)
 \CBox(-12,-12)(12,12){Black}{Yellow}
 \Text(0,0)[]{$\cA^{(1)}_{11}$}
\end{picture}
\end{eqnarray*}
\vskip 0.5cm
\caption{\label{fig:rge2} Graphical representation of the RGE for $n=2$.}
\end{figure}

At $n=3$ we start the exploration of unknown territory -- the RGE read
\bea
\cB_{1}^{(3)}&=&  \cA_{11}^{(3)} \co \nn
\cB_{3}^{(3)}&=& 2\cA_{21}^{(3)} \co \nn
\cB_{3}^{(3)}&=& 3\cA_{31}^{(3)} \co \nn
3 \cA_{33}^{(3)}&=& \nabla_1 \cA_{22}^{(3)} \co \nn
2\cA_{22}^{(3)} + 3\cA_{32}^{(3)}&=& \left(\nabla_1+\nabla_2 \right) 
\left[ \cA_{11}^{(3)}+\cA_{21}^{(3)} \right] \fs
\label{eq:rge_l3_a}
\eea
If we act with $\nabla_1$ on the last equation, we can rewrite the
next-to-last as
\be
3! \cA_{33}^{(3)} = \nabla_1^2 \cA_{11}^{(3)} = \nabla_1^2 \cB_{1}^{(3)} \co
\ee
which shows again that all the information about the three-loop divergences
is contained in the single pole in $\eps$. The equation is however not yet
fully explicit: the operator $\nabla_1$ in fact contains derivatives
$\vec{\partial}^{(n)}$ with all $n$'s, but obviously only the first few may
contribute:
\be
\nabla_1^2 \cB_{1}^{(3)} = \left[\nabla_1^{(1)} \right]^2 \cB_{1}^{(3)} + 
\nabla_1^{(1)} \nabla_1^{(2)} \cB_{1}^{(3)} \co
\ee
where in the last term the operator $\nabla_1^{(1)}$ acts only on
$\vec{\beta}_{1}^{(2)}$ inside $\nabla_1^{(2)}$. From Eq.~(\ref{eq:rge_l2}) we see
that $\nabla_1^{(1)} \cB_{1}^{(2)} = 2 \cA_{22}^{(2)}$, and therefore, that
the HPRGE for $n=3$ can be rewritten as
\be
3! \cA_{33}^{(3)} = (d_1^2+2 d_2) \cB_{1}^{(3)} \fs
\ee

\begin{figure}
\begin{eqnarray}
\nonumber
 3 \qquad
 \begin{picture}(30,40)(0,-2)
 \CBox(-12,-12)(12,12){Black}{Yellow}
 \Text(0,0)[]{$\cA^{(3)}_{33}$}
\end{picture}
& = &
 \begin{picture}(140,40)(-30,-2)
 \CArc(25,0)(25,0,360)
 \CBox(-12,-12)(12,12){Black}{Yellow}
 \Text(0,0)[]{$\cA^{(1)}_{11}$}
 \CBox(38,-12)(62,12){Black}{Yellow}
 \Text(50,0)[]{$\cA^{(1)}_{11}$}
 \Text(75,0)[]{+}
 \CArc(125,0)(25,0,360)
 \CBox(88,-12)(112,12){Black}{Yellow}
 \Text(100,0)[]{$\cA^{(2)}_{22}$}
\end{picture}
\end{eqnarray}
\vskip 0.5cm
\caption{Graphical representation of the RGE for $n=3$.}
\end{figure}

\subsection{Highest-pole equation to all orders}
After having analyzed explicitly the first few cases, it should be now
clear how to extend the derivation of the same equation to all orders. The
only rule that we need to use, is rule n.~3. at the end of
Sect.~\ref{sec:RGE}. The highest pole equation for a generic $n$ reads
\be
n! \cA^{(n)}_{nn} = \nabla_1^{n-1} \cA_{11}^{(n)}  = \nabla_1^{n-1} 
\cB_{1}^{(n)} \fs 
\label{eq:rhp}
\ee
Such a simple expression is suggestive, but not very illuminating, because,
as we have seen above for the $n=3$ case, the product of $\nabla_1$'s is a
complicated object, due to the noncommuting property of the
$\nabla_1$'s. In the $n=3$ case, however, we have seen that one can rewrite
the product of two $\nabla_1$'s in terms of $d_{1,2}$ -- objects which
have a clear meaning and which commute between themselves.
This fact indeed generalizes to all orders, and allows us to rewrite the
product of $n-1$ $\nabla_1$'s (acting on $\cB_{1}^{(n)}$) in terms of
$d_k$'s only with $k \leq n-1$, and to give a clear meaning to the HPRGE:
\be
n \cA_{nn}^{(n)} = \; \left[ \sum_{\underline{\gamma}}
             \frac{1}{\prod_{k=1}^{l}\beta_{k}!}
             (d_{\alpha_{1}})^{\beta_{1}}\cdot ... \cdot
             (d_{\alpha_{l}})^{\beta_{l}} \right] \cB_{1}^{(n)} 
 \label{eq:s_exp}
\ee
where one sums over all $\underline{\gamma}:=\{\alpha_i, \beta_i\}$
having the property $\sum_{i}\alpha_{i}\beta_{i} = n-1$. 
The derivation of this formula can be found in appendix \ref{ap:proof}.

The effect of the operator $d_i$ on any diagram is to substitute a vertex
coming from the Lagrangian $\cL^{(i)}$ with the corresponding highest-pole
counterterm $\cA_{ii}^{(i)}$. The formula (\ref{eq:s_exp}) implies that the
highest pole counterterm is obtained by calculating all one-loop diagrams
contributing to $\cB_{1}^{(n)}$ and substituting {\em all} the vertices from
the Lagrangians $\cL^{(i)}$, $i<n$ with the corresponding highest-pole
counterterm $\cA_{ii}^{(i)}$. 

Notice that all $\cA_{ii}^{(i)}$ can be expressed in terms of $\cB_{1}^{(j)}$'s
with $j \le i$ only: the highest-pole counterterms can all be calculated
with one loop diagrams. On the other hand, Eq.~(\ref{eq:s_exp}) does not
lend itself to an explicit direct evaluation to all orders, but can only be
used recursively. Also, each step up in the recursive procedure is a
nontrivial (although in principle straightforward) one-loop calculation.

\setcounter{equation}{0}
\section{RGE for the subleading poles}
\label{sec:subdiv}

In the case of the subleading poles, the situation becomes somewhat more
complicated. First of all we have to deal with two different possible
$l$-orders for the terms with $k=n-1$. The corresponding equations read
\bea
(n-1) \cA^{(n)}_{n-1 \, n-1} &=& \nabla_1  \cA^{(n)}_{n-2 \, n-2} \nn
n \cA^{(n)}_{n \, n-1} &=& \nabla_1  \cA^{(n)}_{n-1 \, n-2} + \nabla_2
\cA^{(n)}_{n-2 \, n-2} \fs
\label{eq:subdiv1}
\eea
Just like for the highest-pole equation, we need to relate the left-hand side of this
equation with the right-hand side of the equation with $k$ one unit lower, until we
reach $k=1$. One can do this easily for the first equation in
(\ref{eq:subdiv1}) -- the result has exactly the same form as the one for
the highest-pole equation:
\be
(n-1)! \cA^{(n)}_{n-1\, n-1} = \nabla_1^{n-2}  \cA^{(n)}_{11} \fs
\label{eq:sd1}
\ee

In the second equation in (\ref{eq:subdiv1}) the right-hand side contains two terms,
and each of them comes with a different coefficient in the next
equation. The combinatorics is therefore somewhat more complicated, but can
also be worked out without major difficulties.
The result reads
\be
 n! \mA^{(n)}_{n \, n-1} = 2 \nabla_{1}^{n-2}\mA_{21}^{(n)} + \sum_{j=0}^{n-3}
 (n-1-j) \nabla_{1}^{j}\nabla_{2}\nabla_{1}^{n-3-j} \mA_{11}^{(n)} \fs
\label{eq:sd2}
\ee
An analysis of the above formula to all orders is provided in appendix
\ref{sec:bhpe}: the result we obtain is similar to Eq.~(\ref{eq:s_exp}) for
the highest poles, but admittedly considerably more involved. As in the case of
the highest poles, also here the formula can only be used in a recursive
manner: to obtain the subleading poles at \ho $n$ one has first to work out
all leading and subleading poles of \ho $n' < n$, and insert these as
vertices in one- and two-loop diagrams. 
For this reason we provide the discussion of the formula to all orders only
in the appendix and consider here the RGE for the subleading poles for the
first few $\hbar$-orders.

\begin{figure}
\begin{eqnarray}
\nonumber
 2 \qquad
 \begin{picture}(10,60)(0,-2)
 \CBox(-12,-12)(12,12){Black}{Yellow}
 \Text(0,0)[]{$\cA^{(3)}_{22}$}
\end{picture}
& = &
 \begin{picture}(140,60)(-10,-2)
 \CArc(25,0)(25,0,360)
 \CBox(-12,-12)(12,12){Black}{Yellow}
 \Text(0,0)[]{$\cA^{(1)}_{11}$}
 \CBox(38,-12)(62,12){Black}{Yellow}
 \Text(50,0)[]{$\mL^{(1)}$}
 \Text(75,0)[]{+}
 \CArc(125,0)(25,0,360)
 \CBox(88,-12)(112,12){Black}{Yellow}
 \Text(100,0)[]{$\cA^{(2)}_{11}$}
\end{picture} \\
3 \qquad
\begin{picture}(10,60)(0,-2)
 \CBox(-12,-12)(12,12){Black}{Yellow}
 \Text(0,0)[]{$\cA^{(3)}_{32}$}
\end{picture}
& = &
 \nonumber
 \begin{picture}(140,60)(-10,-2)
 \Oval(35,0)(25,35)(360)
 \Line(0,0)(70,0)
 \Vertex(70,0){2}
 \CBox(-12,-12)(12,12){Black}{Yellow}
 \Text(0,0)[]{$\cA^{(1)}_{11}$}
 \Text(80,0)[]{+}
 \CArc(115,0)(25,0,360)
 \CArc(165,0)(25,0,360)
 \CBox(128,-12)(152,12){Black}{Yellow}
 \Text(140,0)[]{$\cA^{(1)}_{11}$}
\end{picture} \\
& + & 
 \nonumber
 \begin{picture}(140,60)(-10,-2)
 \CArc(25,0)(25,0,360)
 \CBox(-12,-12)(12,12){Black}{Yellow}
 \Text(0,0)[]{$\cA^{(2)}_{21}$}
 \end{picture}
\end{eqnarray}
\vskip 0.5cm
\caption{Graphical representation of the RGE for the subleading poles ($n=3$).}
\end{figure}
\subsection{RGE for the subleading poles for $n=2$ and $3$}
The first subleading poles appear at $n=2$. We have already seen the RGE
for this case in the previous section, but we concentrated there on the
leading pole. As for the subleading poles, at this \ho the equations only
relate these to the corresponding $\beta$'s:
\bea
\cB_{1}^{(2)}&=&   \cA_{11}^{(2)}  \co \nn
\cB_{2}^{(2)}&=& 2 \cA_{21}^{(2)}  \fs
\label{eq:subrge_l2}
\eea

At $n=3$ the RGE for the subleading poles start to provide some interesting
information. All the RGE at $n=3$ are given in Eq.~(\ref{eq:rge_l3_a}) --
here we consider only those relevant for the subleading poles, and split
them according to the $l$-order, as in Eq.~(\ref{eq:subdiv1}):
\bea
2 \cA^{(3)}_{22} &=& \nabla_1  \cA^{(3)}_{11} = \left(
  \nabla_1^{(1)}+\nabla_1^{(2)}\right) \cB_{1}^{(3)} \nn 
3 \cA^{(3)}_{32} &=& \nabla_1  \cA^{(3)}_{21} + \nabla_2 \cA^{(3)}_{11}
= {1\over 2} \nabla_1^{(1)} \cB_{2}^{(3)}+ \nabla_2^{(2)} \cB_{1}^{(3)} \fs
\label{eq:subn3}
\eea
The meaning of these equations is as follows: the first one implies that
the part of the subleading poles that comes from two-loop diagrams is
given by the single pole from one-loop diagrams ($\cA^{(3)}_{11}$) after
one has substituted one $\cL^{(1)}$ ($\cL^{(2)}$) vertex with $\cA^{(1)}_{11}$
($\cA^{(2)}_{11}$).
According to the second one the double pole coming from three-loop
diagrams is given by two terms: the first one is obtained from single poles
from two-loop diagrams after substitution of one $\cL^{(1)}$ vertex with
$\cA^{(1)}_{11}$, and the second one from single poles from one-loop diagrams
after substitution of one $\cL^{(2)}$ vertex with $2 \cA^{(2)}_{21}$, i.e. the
subleading poles of one \ho lower.

\begin{figure}
\begin{eqnarray}
 \nonumber
 \begin{picture}(10,60)(0,-2)
 \CBox(-12,-12)(12,12){Black}{Yellow}
 \Text(0,0)[]{$\cA^{(2)}_{11}$}
\end{picture}
& = &
 \begin{picture}(140,60)(-10,-2)
 \CArc(25,0)(25,0,360)
 \CBox(-12,-12)(12,12){Black}{Yellow}
 \Text(0,0)[]{$\mL^{(1)}$}
 \end{picture} \\ 
 \begin{picture}(10,60)(0,-2)
 \CBox(-12,-12)(12,12){Black}{Yellow}
 \Text(0,0)[]{$\cA^{(2)}_{21}$}
\end{picture} 
& = & 
 \nonumber
 \begin{picture}(140,60)(-10,-2)
 \Oval(35,0)(25,35)(360)
 \Line(0,0)(70,0)
 \Vertex(70,0){2}
 \Vertex(0,0){2}
 \Text(80,0)[]{+}
 \CArc(115,0)(25,0,360)
 \CArc(165,0)(25,0,360)
 \Vertex(140,0){2}
\end{picture}
\end{eqnarray}
\vskip 0.5cm
 \caption{Vertices needed for the calculation of the subleading poles ($n=3$).}
\end{figure}

\setcounter{equation}{0}
\section{Role of one particle reducible diagrams} 
\label{sec:eom}

In the graphical representations of the RGE we have always drawn
one-particle-irreducible (1PI) graphs, although also
one-particle-reducible (1PR) ones contribute to the generating functional
$Z$, and possibly to its divergences. It would indeed be desirable to be in
a setting where only 1PI graphs contribute to divergences, and therefore
have to be considered in the RGE. In this section we show how one can
ensure that this setting is realized. The problem is related to the role of
counterterms that vanish at the classical solution of the equations of
motion (EoM terms in short).  At each \ho $n$ all the divergences are
absorbed by the counterterms $\cO_i^{(n)}$, and we may assume that they
form a minimal basis, i.e. that they are all linearly independent. In the
reduction procedure of a complete list of counterterms to a minimal basis
one eliminates terms which are algebraically linearly dependent (such as
terms related by trace identities in CHPT), terms which are a total
derivative, and also the EoM terms. While the first two categories of terms
can be eliminated without any consequences on the renormalization
procedure, eliminating EoM terms is a less trivial issue, which has to be
discussed in some detail. This problem has already been dealt with in
Refs.~\cite{Ecker:1996rk,Bijnens:1998yu,Bijnens:1999hw}, in performing the
renormalization at the two-loop level. As observed (and explicitly
verified) in \cite{Bijnens:1998yu}, one can choose the coefficient in front
of the EoM terms in such a way that the sum of 1PR graphs is finite. Here we
discuss how this can be done to any loop order.

The sum of all 1PI graphs defines the generating functional of proper
vertices (or effective action) $\Gamma$ -- we denote the sum of 1PR
graphs with $Z^\mathrm{1PR}$. In the background field method one shifts
the fields over which the path integral is performed, $\phi \to
\phi+\xi$, and integrates over the $\xi$ fields. In this framework
$\Gamma$, the sum of all 1PI diagrams, becomes a functional both of the
fields $\phi$ (which need not be fixed at the solution of the EoM) and of
the external sources $J$. Order by order in the loop expansion we have 
\be
Z_n[J]=\Gamma_n[\phi,J]_{|_{\phi=\phi_{cl}}}+Z_n^\mathrm{1PR}[J] \co
\ee
where the contribution of 1PI diagrams to $Z_n$ is obtained by evaluating
$\Gamma_n$ at the classical solution $\phi_{cl} =\phi_{cl}[J]$. In the
following we will denote with a bar a functional which is evaluated at the
classical solution: $\bar \Gamma := \Gamma_{|_{\phi=\phi_{cl}}}$.
We stress that the splitting of $Z_n[J]$ between 1PI and
1PR diagrams is ambiguous: either by a field redefinition, or by adding terms
that vanish at the EoM (the two things are equivalent, see, e.g. 
\cite{Bijnens:1999sh}) one can change $\Gamma[\phi,J]$.  
On the other hand, if all counterterm Lagrangians $\mL^{(k)}$ for all 
$k \leq n$ have been specified, including terms that vanish at the EoM,
then $\Gamma[\phi,J]$ is unambiguously defined.

The most important property of the effective action $\Gamma$ in this
context is that $Z_n^\mathrm{1PR}$ can be written as tree diagrams with
$\Gamma_k$'s (with $k < n$) as vertices \cite{WeinbergQFT,PeskinQFT}.  At
the two- and three-loop level, e.g., we have 
\bea
Z_2^\mathrm{1PR}&=& -{1 \over 2} \bar \Gamma_1^i G_{il} \bar \Gamma_1^l \nn
Z_3^\mathrm{1PR}&=& {1 \over 2} \bar \Gamma_1^i G_{ij} \bar \Gamma_1^{jk}G_{kl}
\bar \Gamma_1^l- \bar \Gamma_2^i G_{ij} \bar \Gamma_1^j \co
\label{eq:Z31PR}
\eea
where we have denoted functional derivatives with respect to $\phi$ with
\be
\Gamma_n^{i_1 \ldots i_k}:= { \delta^k \Gamma_n \over \delta \phi_{i_1}
  \ldots \delta \phi_{i_k}} \fs
\label{eq:fder}
\ee
In general all vertices of the form $\Gamma_k^{j_1 \ldots j_m}$ with $k <
n\; ;  \; j_m \leq n-k $ contribute to $Z_n^\mathrm{1PR}$. 

The condition that ensures that both $\bar \Gamma_n$ and $
Z_n^\mathrm{1PR}$ are separately finite can be established by
induction. Suppose that for all $k<n$ $\bar \Gamma_k$ and 
$Z_k^\mathrm{1PR}$ are separately finite.
Since tree diagrams do not generate new divergences, $Z_n^\mathrm{1PR}$ can
be divergent only if some of the vertices are: in order to have
$\bar Z_n^\mathrm{1PR}$ finite we must impose that all the vertices
$\bar \Gamma_k^{j_1 \ldots j_m}$ with $k<n$ and $m \leq n-k$ are finite. 
But all $\Gamma_k$ for $k<n$ are finite at the EoM by assumption:
possible divergences in their functional derivatives can be described by
local terms that vanish at the EoM. These can be removed by tuning the
coefficients of the counterterms of \ho $k$ that vanish at the EoM. 
The proof by induction is completed by observing that for $n=0$ and $1$ the
generating functional does not admit divergent 1PR contributions: $\bar
\Gamma_0$ and $\bar \Gamma_1$ are both finite (after renormalization). A
detailed  discussion of the case $n=2$ can be found in
Ref.~\cite{Bijnens:1998yu}. 

In summary, in order to have 1PR and 1PI contributions to the generating
functional $Z$ separately finite, one must renormalize not only the $Z[J]$
but also the effective action $\Gamma[\phi,J]$ as a functional of
$\phi$. This can be done by exploiting the freedom to add counterterms that
vanish at the EoM.

What we are interested in here, however, is not whether $Z_n^\mathrm{1PR}$
is completely finite, but whether it plays a role in the RGE, and in
particular in the HPRGE. According to the RGE the highest poles
are fully determined by the $1/\eps$ divergences of \lo one: in
order not to have to consider 1PR graphs in the HPRGE, we must impose that
$Z_n^\mathrm{1PR}$ does not contain $1/\eps$ divergences of \lo one. In
view of the general statements made above about $Z_n^\mathrm{1PR}$ this
condition has to be transferred to the vertices $\Gamma_k$'s for all $k
\leq n-1$. The part of $\Gamma_k$ which is of \lo one can be
projected out with the operator:
\be
P_{l=1}^k:= \sum_{\alpha_{i}\beta_{i} = k-1} \prod_{i} {1 \over \beta_i!}
\left[ D^{(\alpha_i)} \right]^{\beta_{i}} \co
\label{eq:Pl1}
\ee
where $[D^{(i)}]^1=D^{(i)}$ and $\left[ D^{(i)} \right]^{n+1}:=
\left( D^{(i)}-n \right) \left[D^{(i)}\right]^n$, cf. Eq.~(\ref{eq:nv}).
The condition we have to impose at order $n$ therefore reads
\be
\left[ P_{l=1}^k \Gamma_k \right]^{j_1 \ldots j_m} = \mbox{finite} 
\label{eq:eom_cond}
\ee
for all $k <n$ and $m \leq n-k$.
At first sight, this may look like a severe complication of the
implementation of the RGE. As we discuss in appendix~\ref{ap:gammas},
however, this is not the case: typically, for a given $\Gamma_k$ only a 
finite number of functional derivatives may at all be divergent, such that 
whenever a new $\Gamma_k$ is calculated and renormalized, one can impose
that all its functional derivatives be finite (by an appropriate choice of
the counterterm basis). If one renormalizes the theory in this manner the
1PR diagrams play no role in the renormalization procedure, and therefore
in the RGE. An illustration of the concepts discussed here is provided in
the following section for the case of renormalizable theories and in
appendix~\ref{ap:gammas} in the general case.

\setcounter{equation}{0}
\section{RGE for renormalizable theories}
\label{sec:ren}
The RGE we discussed so far were derived in a framework which is
particularly convenient with nonrenormalizable theories like CHPT. This
setting, on the other hand, is completely general, and can be used also
with renormalizable theories, as we want to show in this section.  For the
sake of simplicity we will discuss the case of a $O(N)$ $\phi^4$ theory.

The $O(N)$ $\phi^4$ theory is defined by the following classical Lagrangian:
\be
\cL^{(0)}={1 \over 2} \left( \partial_\mu \vec{\phi} \cdot \partial^\mu
  \vec{\phi} -M^2 \vec{\phi}\cdot \vec{\phi} \right) - {\lambda \over 4} 
\left( \vec{\phi}\cdot \vec{\phi} \right)^2 -\vec{\phi} \cdot \vec{f}
\co 
\label{eq:Lphi4}
\ee
where $\vec{\phi}$ and $\vec{f}$ are $N$-component vectors, the latter
of external fields. This theory is renormalizable: the divergences arising
in loop calculations can be reabsorbed by a redefinition of the wave
function renormalization $Z_\phi$ and the bare parameters $M^2$ and
$\lambda$. Here, however, we want to discuss renormalization in a manner
which is completely analogous to the case of a nonrenormalizable theory. We
will introduce a new Lagrangian at each order in $\hbar$:
\be
\cL^{\rm bare}=\cL^{(0)}+\hbar \cL^{(1)}+\hbar^2 \cL^{(2)} + \ldots \co
\ee
with
\be
\cL^{(n)} = c^{(n)}_1 {1 \over 2} \partial_\mu \vec{\phi} \cdot
\partial^\mu \vec{\phi} -  c^{(n)}_2 {1 \over 2} M^2 \vec{\phi}\cdot \vec{\phi} 
- c^{(n)}_3 {\lambda \over 4}\left(\vec{\phi}\cdot \vec{\phi} \right)^2
\fs
\ee
Notice that by using the equations of motion one can eliminate one of the
four possible terms \eqref{eq:Lphi4} such that a minimal basis of
counterterms counts only three independent operators.

In order to renormalize the theory it is sufficient to define the bare
couplings $c^{(n)}_i$ as
\be
c^{(n)}_i = \left(\mu^{-\eps} \lambda \right)^n \left[c^{(n)\, r}_i+
  \sum_{k=1}^n a^{(n)}_{k \,i} \eps^{-k} \right] \co
\label{eq:bareC}
\ee
where, for convenience we have factored out $\mu^{-\eps n} \lambda^n$ in
such a way that the renormalized couplings $c^{(n)\, r}_i$  and the
coefficients $a^{(n)}_{k \,i}$ are dimensionless even in $d \neq 4$. With
this choice the scaling with $\mu$ of the bare Lagrangians $\cL^{(n)}$ is
exactly as given in Eq.~\eqref{eq:Lbare}: the RGE that follow from there
must therefore be valid also in this case.
The coefficients $a^{(n)}_{k \,i}$ are analogous to those defined in
\eqref{eq:Ank}, and can further be expanded in the $l$-order: $a^{(n)}_{k\,i}
= \sum_{l=k}^n a^{(n)}_{lk\,i}$. Each new coupling constant $c^{(n)\, r}_i$
has its own beta function, defined as
\be
\mu {d \over d \mu} c^{(n)\, r}_i(\mu) = \beta^{(n)}_i + \eps n c_i^{(n)}
\co 
\ee
and which the RGE relate to
\be
\beta^{(n)}_i = \sum_{l=1}^n l a_{l1 \, i}^{(n)} \fs
\ee

A full two-loop calculation of the generating functional in this theory is
relatively easy -- the results can be found, e.g., in
\cite{Bijnens:1997vq} and read:
\be
a^{(1)}_{1 \,1}=0 \quad
a^{(1)}_{1 \,2}={2 \over P} (N+2) \quad
a^{(1)}_{1 \,3}={2 \over P} (N+8) \co
\label{eq:renphi4_1}
\ee
for the one-loop divergences and
\bea
a^{(2)}_{1 \,1}&=&-{1 \over P^2} (N+2) \nn a^{(2)}_{2 \,1}&=&0 \nn 
a^{(2)}_{1 \,2}&=&-{2 \over P^2 }(N+2) \left[3-P(c^{(1)\, r}_2 + c^{(1)\,
  r}_{3} )\right] \nn 
a^{(2)}_{2 \,2}&=&{4 \over P^2 } (N+2)(N+5) \nn
a^{(2)}_{1 \,3}&=&-{4 \over P^2}\left[22+5N-P(N+8)c^{(1)\, r}_{3}\right]\nn 
a^{(2)}_{2 \,3}&=&{4 \over P^2 } (N+8)^2 
\label{eq:renphi4_2}
\eea
at two loops, with $P=16 \pi^2$. The results of the two-loop divergences
allow us to test the RGE for $n=2$ which are written explicitly in
\eqref{eq:rge_l2}: 
\bea
2 a^{(2)}_{2 \,1}&=& \nabla a^{(2)}_{1 \,1} = \sum_{i=1}^3 a^{(1)}_{1 \,i}
{\partial \over \partial c^{(1)\, r}_i} a^{(2)}_{1 \,1} = 0 \co \nn
2 a^{(2)}_{2 \,2}&=&\nabla a^{(2)}_{1 \,2} = \sum_{i=1}^3 a^{(1)}_{1 \,i}
{\partial \over \partial c^{(1)\, r}_i} a^{(2)}_{1 \,2} ={4 \over P^2 }
(N+2)(2N+10) \co \nn 
2 a^{(2)}_{2 \,3}&=& \nabla a^{(2)}_{1 \,3} = \sum_{i=1}^3 a^{(1)}_{1 \,i}
{\partial \over \partial c^{(1)\, r}_i} a^{(2)}_{1 \,3} ={8 \over P^2 }
(N+8)^2 \co
\eea
which perfectly agree with the results of the direct calculation
\eqref{eq:renphi4_2}. We can now extend and solve the RGE for the highest
poles to all orders. What makes such a solution possible in a
renormalizable theory is the fact that the new vertices coming from the
$\cL^{(n)}$ Lagrangian (and that have to be inserted in the relevant
one-loop diagrams) are identical to those appearing in the classical
Lagrangian \eqref{eq:Lphi4}.

A divergent one-loop contribution to the renormalization of $c^{(n)}_3$ 
can have at most two vertices, i.e. must be proportional to $c^{(n-k) \, r}_3
c^{(k-1) \, r}_3 $, with $k=1\ldots n$. The coefficient of all these terms
is identical, and can be read off from \eqref{eq:renphi4_1}:
\be
a^{(n)}_{1 1 \, 3} ={2 \over P} (N+8) \sum_{k=1}^n c^{(n-k) \, r}_3 c^{(k-1)
  \, r}_3 \fs
\label{eq:an13}
\ee
The equation to all orders \eqref{eq:s_exp} can now be easily solved by
induction. We have seen that
\be
a^{(1)}_{1 \, 3} = p \qquad a^{(2)}_{2 \, 3} = p^2 \qquad 
p:={2(N+8) \over P}  \co
\ee
and it is easy to prove that if $a^{(n-1)}_{n-1 \, 3}=p^{n-1}$, then
Eq.~\eqref{eq:an13} and \eqref{eq:s_exp} imply that 
$a^{(n)}_{n \, 3}=p^n$. 

The solution of the RGE for the mass term is only slightly more
complicated, but can also be solved to all orders. 
The divergent part of all one-loop graphs at \ho $n$ can also in this case
depend on at most two counterterms, $c^{(k) \, r}_3 c^{(n-1-k) \, r}_2$,
for all $0 \leq k \leq n-1$. By exploiting again the fact that the
structure of the counterterms is identical to that of the classical
Lagrangian we can read off from \eqref{eq:renphi4_1} the coefficient of all
the divergent terms:
\be
a^{(n)}_{1 1 \, 2} = {q \over n}\sum_{k=0}^{n-1} c^{(k) \, r}_3 
c^{(n-1-k) \, r}_2 \qquad q:={2 \over P} (N+2) \fs
\label{eq:an12}
\ee

At one and two loops we had 
\be
a^{(1)}_{1 \, 2} = q \qquad a^{(2)}_{2 \, 2} = {1 \over 2} q (q+p) \co
\ee
and by using Eq.~\eqref{eq:an12} and \eqref{eq:s_exp} we can easily prove
by induction that, if 
\be
a^{(n-1)}_{n-1 \, 2} = {1 \over (n-1)!} \prod_{k=0}^{n-2} (q+kp) \co
\ee
then
\be
a^{(n)}_{n \, 2} = {1 \over n!} \prod_{k=0}^{n-1} (q+kp) \fs
\ee

\subsection{EoM counterterms}
The example of the $O(N)$ $\phi^4$ Lagrangian is useful also to illustrate
what is the role of EoM counterterms. For the $\cL^{(n)}$ Lagrangians we
chose above the set of operators 
\be
\cO_1={1 \over 2} \partial_\mu \vec{\phi} \cdot
\partial^\mu \vec{\phi} \co \qquad
\cO_2=-{1 \over 2} M^2 \vec{\phi}\cdot \vec{\phi} \co \qquad
\cO_3=-{\lambda \over 4}\left(\vec{\phi}\cdot \vec{\phi} \right)^2 \co
\label{eq:Olist}
\ee
but we could in principle choose a different set at each order $n$,
and replace one of the operators in \eqref{eq:Olist} with
\be
\cO_4=-\vec{f}\cdot \vec{\phi} \fs
\ee

The general discussion given in Sect.~\ref{sec:eom} shows that if one wants
to deal only with 1PI diagrams in the RGE, one should select the counterterm
Lagrangian requiring that even functional derivatives with respect to
$\phi$ of the effective action $\Gamma_n$ be finite.
In the present case, at one loop we have:
\bea
\Gamma_1&=&S_1+{1 \over 2} \mbox{Tr} \log S_0^{ij} \co \nn
\Gamma_1^k&=&S_1^k+{1 \over 2} S_0^{ijk} G_{ij} \fs
\label{eq:gamma1}
\eea
The evaluation of the loop part of $\Gamma_1^k$ gives
\be
{1 \over 2} S_0^{ijk}G_{jk} =-{ 1 \over \eps}{2 \over P }
\left\{\phi_i \left[M^2(N+2)+\lambda \vec{\phi}\cdot \vec{\phi} (N+8)
  \right] \right\} + \mbox{finite terms} \co
\ee
while the functional derivative of the four operators entering
at each order reads
\be
{\delta \cO_1 \over \delta \phi_i}  = - \partial^2 \phi_i \co \quad
{\delta \cO_2 \over \delta \phi_i}  = -M^2 \phi_i \co \quad
{\delta \cO_3 \over \delta \phi_i}  = -\lambda \phi_i 
\left( \vec{\phi} \cdot \vec{\phi} \right) \co \quad
{\delta \cO_4 \over \delta \phi_i}  =-f_i \co
\ee
which shows that the divergent part in $ S_0^{ijk}G_{jk}/2$ can be
cancelled by $\bar S_1^{i}$ only if both $\cO_2$ and $\cO_3$ are included
in the list of operators: at one loop either $\cO_1$ or $\cO_4$ can be
eliminated. In fact, since the structure of the Lagrangian is always the
same, this criterion extends also to higher orders.

In this example it is also easy to verify (and we leave this to the reader)
that in case one choses a different basis for the counterterm Lagrangian,
e.g. $\cO_1$, $\cO_2$ and $\cO_4$, then 1PR diagrams do contribute to local
divergences, and that if one takes them also into account then the RGE
still hold as before.  Fixing the basis in such a way that only 1PI graphs
contribute to the divergences is only a matter of convenience, and does not
change the form of the RGE.

\subsection{Comparison to the standard treatment of $\lambda \phi^4$}
The way we discussed renormalization for this example of a renormalizable
theory is not standard, as it is designed to parallel the renormalization
procedure in CHPT (or any other nonrenormalizable theory). In this section
we will clarify the connection between our treatment and the standard one.

The multiplicity of coupling constants we introduced, the $c_i^{(n)}$, are
not separately observable. Indeed, the physical mass and coupling constant
that will appear in any observable will have the form
\bea
M^2_{\rm ph}&=& 
M^2\left(1+\sum_{n=1}^\infty a^{(n)}_{n \,2} \ell^n + \ldots
  +\sum_{n=1}^\infty \lambda^n c^{(n)\,r}_{2}(\mu) \right) 
\simeq M^2\left(1-p \ell \right)^{-q/p} \co \nn
\lambda_{\rm ph}&=&\lambda \left(1+ \sum_{n=1}^\infty a^{(n)}_{n \,3} 
\ell^n + \ldots +\sum_{n=1}^\infty \lambda^n c^{(n)\,r}_{3}(\mu) 
\right)\simeq {\lambda \over 1-p\ell} \co 
\label{eq:Mph}
\eea
where $\ell=\lambda \log \mu$, and where the ellipses denote terms with
subdominant powers of logs, and where the last expression is accurate only
up to the leading logs. Notice that the expression between brackets
after the first equality sign in \eqref{eq:Mph} is $\mu$-independent:
$\lambda$ and $M$ are $\mu$-independent by definition, and the
coefficients of each power of $\lambda$ are separately $\mu$-independent
as implied by the $\beta$-functions of the $c^{(n)\,r}_{i}$ couplings.

Since $\lambda$ and $c^{(i) \, r}_3(\mu)$ always appear in this
combination, and are not separately observable, it is useful to lump them
together into one single quantity, the part in $\lambda_{\rm ph}$ which
does not contain logs:
\be
\lambda_R(\mu):=\lambda \left(1+\sum_{i=1}^\infty \lambda^i 
c^{(i) \, r}_3 (\mu) \right) \co
\label{eq:lambdaR}
\ee
and which is nothing but the standardly defined renormalized coupling
constant. Any quantity can now be expressed in terms of $\lambda_R(\mu)$
(and the similarly defined renormalized mass), rather than $\lambda$, by
inverting \eqref{eq:lambdaR}. 

It is interesting to derive how $\lambda_R(\mu)$ depends on $\mu$ from
Eq.~\eqref{eq:lambdaR}: 
\be
\beta_\lambda:= \mu{d \over d \mu} \lambda_R(\mu) = \nabla 
\lambda_R(\mu) = \lambda \sum_{i=1}^\infty 
\lambda^i \beta^{(i)}_3  \co 
\ee
and if we reexpress $\lambda$ on the r.h.s in terms of $\lambda_R(\mu)$ we
finally get
\be
\beta_\lambda = \lambda_R(\mu) \sum_{i=1}^\infty 
\lambda^i_R(\mu) \bar \beta^{(i)}_3 \co
\ee
where
\be
\bar \beta^{(i)}_3=\left[\beta^{(i)}_3\right]_{|_{c_i^{(n)}=0}} \co
\ee
is the part of the beta functions which does not depend on any of the
constants. (Which can also be identified with the help of the loop number:
$\beta^{(i)}_3=\sum_l \beta^{(i)}_{l \,3}$; $\bar
\beta^{(i)}_3=\beta^{(i)}_{i \,3}$.)

\setcounter{equation}{0}
\section{RGE for theories with renormalizable and nonrenormalizable
interactions} 
\label{sec:phi6}
In the previous section we have shown how the RGE derived here apply to the
case of a renormalizable field theory. Our whole framework, on the other
hand, has been developed in order to treat the case of a nonrenormalizable
field theory like CHPT, where the only renormalizable part of the
Lagrangian corresponds to a free field theory. Between these two extreme
cases there is an intermediate one, which is actually used quite often in
phenomenology: the case of an interacting renormalizable theory to which
one adds nonrenormalizable interactions. The latter are usually suppressed
by powers of a certain energy scale.
For example the search for new physics beyond the Standard Model is often
performed in such a framework -- translating experimental measurements into
bounds on the strength of the nonrenormalizable interactions.
In the present section we will discuss the application of our RGE to such
cases. The role of EoM terms in such a framework has been discussed in
\cite{Einhorn:2001kj}. As in the previous section we choose to work with a
simple example and illustrate in the clearest possible setting the use of
the RGE. We take as renormalizable part of the Lagrangian the $O(N)$
$\phi^4$ theory discussed above and add a $\phi^6$ term to it:
\be
\cL^{(0)}={1 \over 2} \left( \partial_\mu \vec{\phi} \cdot \partial^\mu
  \vec{\phi} -M^2 \vec{\phi}\cdot \vec{\phi} \right) - {\lambda \over 4} 
\left( \vec{\phi}\cdot \vec{\phi} \right)^2 - {g \over 6 \Lambda^2 } 
\left( \vec{\phi}\cdot \vec{\phi} \right)^3 -\vec{\phi} \cdot \vec{f}
\label{eq:Lphi6}
\co
\ee
where $g$ is a dimensionless coupling constant and $\Lambda$ an energy
scale. Such a Lagrangian is nonrenormalizable, i.e. more and more
counterterms will be required in order to make loop calculations finite. We
choose, however, to neglect all the effects which are suppressed by more
powers of $\Lambda$ than the $\phi^6$ interaction. With this choice, even
at higher orders we will consider only the four operators introduced in
(\ref{eq:Lphi6}):
\be
\cL^{(n)} = c^{(n)}_1 {1 \over 2} \partial_\mu \vec{\phi} \cdot
\partial^\mu \vec{\phi} - c^{(n)}_2 {1 \over 2} M^2 \vec{\phi}\cdot \vec{\phi} 
- c^{(n)}_3 {\lambda \over 4}\left(\vec{\phi}\cdot \vec{\phi} \right)^2
- c^{(n)}_4 {g \over 6 \Lambda^2}\left(\vec{\phi}\cdot \vec{\phi} \right)^3
\fs
\ee
If we define the bare couplings as in Eq.~(\ref{eq:bareC}) and renormalize
the theory at the one-loop level we find the following nonzero coefficients
\bea
a^{(1)}_{1 \,2}&=&{2 \over P} (N+2) \nn
a^{(1)}_{1 \,3}&=&{2 \over P} \left[ (N+8) + 2 \eta(N+4) \right] \nn
a^{(1)}_{1 \,4}&=&{6 \over P} (N+14) \co
\eea
where
\be
\eta:= {g M^2 \over \lambda^2 \Lambda^2} \fs
\ee
The RGE allow us to move to higher loops: at the two-loop level,
e.g. we find
\bea
a^{(2)}_{2 \,2}&=&{1 \over P} (N+2) \left[a^{(1)}_{1 \,2}+a^{(1)}_{1 \,3}
\right] \nn
\tilde a^{(2)}_{2 \,3}&=&{2 \over P} \left[ (N+8) a^{(1)}_{1 \,3} + \eta(N+4) 
\left(a^{(1)}_{1 \,2}+a^{(1)}_{1 \,4} \right)\right] \nn
a^{(2)}_{2 \,4}&=&{3 \over P} (N+14) \left(a^{(1)}_{1 \,3}+a^{(1)}_{1 \,4}
\right) \co 
\eea
where, for convenience, we have introduced the symbol
\be
\tilde a^{(n)}_{n \,3}=a^{(n)}_{n \,3}+\eta \hat a^{(n)}_{n \,3}
\ee
which explicitly shows that for $\eta=0$ we obtain exactly the same
coefficients $a^{(n)}_{n \,3}$ as in the renormalizable case.
If we apply the RGE to higher orders following the same reasoning used in
the previous section we get the general results
\bea
\hat a^{(n)}_{n \,3} &=&{1 \over n} {4 \over P} (N+4) \sum_{k=1}^n 
a^{(n-k)}_{n-k \; 2} a^{(k-1)}_{k-1 \; 4} \nn
a^{(n)}_{n \,4}&=&{1 \over n} {6 \over P} (N+14) \sum_{k=1}^n
a^{(n-k)}_{n-k \; 3} a^{(k-1)}_{k-1 \; 4}  \co 
\eea
where we have considered only the new parts with respect to the purely
renormalizable case. These recursion relations can be solved also in this
case and give
\be
\hat a^{(n)}_{n \,3}= {r \over n!} \prod_{k=0}^{n-1} (t+kp) \; ,
\qquad
a^{(n)}_{n \,4}= {1 \over n!} \prod_{k=0}^{n-1} (s+kp) \co
\ee
where $p$ has been introduced in the previous section and
\be
r:={4(N+4) \over P} \co \qquad s :={6(N+14)  \over P} \co \qquad
t:=q+s= {8(N+11) \over P} \fs
\ee

In our formulation of the renormalization procedure an infinite number of
finite counterterms appear, which are not individually observable, as
already discussed for the $\phi^4$ theory. Only the sum over the series of
counterterms and the accompanying logs are observable quantities. For the
coupling constant of the new $\phi^6$ interaction we find
\bea
g_\mathrm{ph}&=& g \left(1+ \sum_{n=1}^\infty a^{(n)}_{n \,4} 
\ell^n + \ldots +\sum_{n=1}^\infty \lambda^n c^{(n)\,r}_{4}(\mu) 
\right) + O\left( {g^2 M^2 \over \Lambda^2} \right) \nn
 &\simeq& g(1-\ell p)^{-s/p} + O\left( {g^2 M^2 \over \Lambda^2} \right) \co
\eea
which shows that even to first order in $g$ the observable coupling
constant gets renormalized in a nontrivial way by the renormalizable part
of the interaction, and that the corresponding series of leading logs can
be resummed. The results obtained here are in agreement with what one would
obtain by considering the standard treatment of the scaling behaviour of
operators -- the solution of our general RGE's
in this particular case has shown that the series of the leading logs is
determined by one single parameter, $s$, which is nothing but the anomalous
dimension (modulo normalization factors) of the $\phi^6$ operator.

As far as the renormalizable $\phi^4$ interaction is concerned, we
have seen that the $\phi^6$ term renormalizes it at every order in
the loop expansion. The renormalization is proportional to $gM^2/\Lambda^2$
and requires the introduction of a specific counterterm: the one
for the $\phi^4$ term now must have the form
\be
\tilde c_3^{(n)}=c_3^{(n)} + {gM^2 \over \Lambda^2} \hat c_3^{(n)} \co
\ee
where both $c_3^{(n)}$ and $\hat c_3^{(n)}$ have to be split into infinite and
finite, scale-dependent parts. The finite, observable coupling constant
now becomes
\bea
\lambda_\mathrm{ph}&=&\lambda \left(1+ \sum_{n=1}^\infty a^{(n)}_{n \,3}
\ell^n + \ldots +\sum_{n=1}^\infty \lambda^n c^{(n)\,r}_{3}(\mu) 
\right) \nn
&+& r {gM^2 \over \Lambda^2} \left(1+ \sum_{n=1}^\infty \hat a^{(n)}_{n\,3}
\ell^n + \ldots + +\sum_{n=1}^\infty \lambda^n \hat c^{(n)\,r}_{3}(\mu) 
\right) \fs
\label{eq:laph}
\eea
At first sight one may get the impression that, if we were now to resum the
series of the leading logs in $\lambda_\mathrm{ph}$, and define the
corresponding $\lambda_R(\mu)$, this would scale differently from the
renormalizable case. However we first notice that the correction
proportional to $gM^2/\Lambda^2$ in Eq.~(\ref{eq:laph}) is scale independent. 
Indeed one can easily check that the series of the leading logs in the
second term on the right-hand side of Eq.~(\ref{eq:laph}) can be fully
reabsorbed by substituting $gM^2/\Lambda^2 \to
g_\mathrm{ph}M_\mathrm{ph}^2/\Lambda^2$. The leading log approximation for
$\lambda_\mathrm{ph}$ therefore reads
\be
\lambda_{\rm ph} \simeq {\lambda \over 1-p\ell} + r { g_\mathrm{ph}
  M^2_\mathrm{ph} \over \Lambda^2} + O(g^2) \simeq {\tilde \lambda \over 1-p\ell}
\ee
with $\tilde \lambda = \lambda+ r g_\mathrm{ph} M^2_\mathrm{ph} /
\Lambda^2$. The conclusion is that even in the presence of a $\phi^6$
interaction the scaling behaviour of the $\phi^4$ coupling constant does
not change, provided one uses the same renormalization condition for
$\lambda$.

\setcounter{equation}{0}
\section{Conclusions}
\label{sec:conc}
In this paper we have studied the RGE for a generic nonrenormalizable
QFT. In the formulation of the problem we have adopted a notation suited to
the case of CHPT, but have not used any of its specific properties in the
derivation of the equations: the RGE that we derived are completely general.

We have worked out explicitly the structure of the leading divergences to
all orders, and found out that they can be recursively expressed in terms
of divergences of one-loop diagrams only. This result is an extension to
all orders of the result obtained by Weinberg at the two-loop level
\cite{Weinberg:1979kz}. Like in that case, however, where in order to
obtain the leading two-loop divergence one had to perform a new and
nontrivial one-loop calculation \cite{Colangelo:1995jm,Bijnens:1998yu},
the extension to higher loops also requires at each step a new one-loop
calculation. In the case of CHPT, e.g., such one-loop calculations are,
although straightforward in principle, long and tedious in practice. As we
do not know a way to perform all these calculations in one go, and solve
explicitly the recursive procedure, we are not able to provide a method to
make resummations of series of leading chiral logs.

A technical problem which occurs in the practical use of the RGE concerns
the role of 1PR diagrams. This in turn is related to the freedom one has in
choosing a basis for the counterterms at each order in the perturbative
expansion, and to the fact that different bases may be related by
counterterms that vanish at the solution of the equations of motion.
As we have shown one can use this freedom to choose the basis at each order
in such a way that in the RGE only 1PI one-loop diagrams have to be
considered. Alternatively, if one wants to use an arbitrary basis then the
RGE provide the right answer for the leading divergences only if one takes
into account also the local divergences generated by 1PR one-loop graphs.

We have analyzed also the RGE for the subleading divergences, but there
even a fully explicit recursive relation is too complicated to write
down. We have discussed explicitly the equations at the two- and
three-loop level. A discussion of how one can derive the all-order
formula can be found in appendix.

If one formulates the renormalization procedure for a renormalizable QFT by
introducing at each loop order (or order in $\hbar$) a new bare Lagrangian
which is independently scale invariant, the RGE which we have derived here
apply equally well to this case. We have shown in the explicit example of a
$\phi^4$ theory that one can solve explicitly the recursion relations for
the leading divergences and obtain results which are in full agreement with
those obtained in the usual formulation of renormalizable QFT. We could
calculate explicitly the series of the leading divergences even after
adding a $\phi^6$ interaction to the $\phi^4$ theory: in this case the
results provided a calculation of the anomalous dimension of the $\phi^6$
operator. These explicit examples illustrate neatly why the complicated
structure of the RGE that we have derived becomes manageable for the case
of a renormalizable theory: the structure of the counterterm
Lagrangian is the same to all orders in $\hbar$, and this makes the
solution of the recursion relations possible. 

This is unfortunately not the case for nonrenormalizable theories of the
CHPT kind, where a resummation of the leading divergences does not seem to
be feasible. In the past, applications of the RGE in CHPT have concerned
the calculations of double chiral logs for various quantities. We plan to
extend these calculations to other quantities of interest, namely weak
nonleptonic decays, where these double chiral logs will provide interesting
informations about the next-to-next-to-leading corrections. It will also be
very interesting to explore the practical feasibility of calculations of
triple chiral logs for some simple quantities, e.g. $M_\pi$, and see how
far one can push the calculation of the series of the leading logs.

\section*{Acknowledgements}
We are indebted to B.~Ananthanarayan for asking questions about the use of
renormalization group equations in CHPT that prompted the present analysis,
and for collaboration at an early stage of this work. We thank J\"urg
Gasser, Peter Hasenfratz and Daniel Wyler for discussions, and Heiri
Leutwyler for discussions and comments on the manuscript. This work was
supported by the Swiss National Science Foundation and by the TMR,
EC-Contract HPRN--CT2002--00311 (EURIDICE).

\appendix

\renewcommand{\theequation}{\thesection.\arabic{equation}}
\setcounter{equation}{0}
\section{Dimensional analysis}
\label{ap:diman}
In deriving the RGE we have used as starting point the scaling with $\mu$
of the Lagrangians of \ho $n$ given in Eq.~(\ref{eq:Lbare}), and have
justified this choice with the claim that in CHPT this is the correct
one. We explain this here. The leading order CHPT Lagrangian (for
simplicity we work here in the chiral limit) reads \cite{Gasser:1984yg} 
\be
\mL^{(0)}={F^2 \over 4} \langle \partial_\mu U^\dagger \partial^\mu U \rangle
\ee
with $U=\exp i\phi/F$ a dimensionless function of $\phi$, which implies
$[\phi] = [F]$.  The dimension of $\mL^{(0)}$ (which is $\mu$-independent
by definition) is $d$, which implies $d = 2[\partial_{\mu}]+2[F]$. This
leads to $[\phi] = [F] = \frac{d-2}{2}$.

At higher orders the Lagrangians $\mL^{(n)}$ will contain $2(n+1)$ powers
of derivatives which, at $d=4$ must be compensated by $2(n+1)-4=2(n-1)$
inverse powers of a physical energy scale. The only available one in this
framework is $F$.  The correct dimensions of $\mL^{(n)}$ for $d \neq 4$ can
be restored by the appropriate powers of the arbitrary scale $\mu$.  The
Lagrangian $\mL^{(n)}$ must therefore scale with a factor $ \mu^{- \eps n}$.

\setcounter{equation}{0}
\section{Proof of the highest pole equation to all orders}
\label{ap:proof}

\subsection{Notation}

In the RGE products of $\nabla_l$'s appear everywhere, and in
order to fully exploit the information contained in the RGE it
is necessary to express them explicitly. In this appendix we show
how to do this. We first introduce some convenient notation, and denote a
product of $k$ $\nabla_{l_i}$'s simply by the (ordered) list of $l_i$'s
within square brackets:
\be
\nabla_{l_1} \nabla_{l_2} \ldots \nabla_{l_k} =: [ l_1 l_2 \ldots l_k] \fs 
\ee
We remind the reader that the subscripts $l_i$'s stand for the $l$-order
of the $\beta$-functions appearing inside the $\nabla$'s.
Such a product contains many terms because each $\nabla_{l_i}$ can act on
all other $\nabla_{l_j}$ on its right-hand side. In order to handle these
many terms conveniently it is necessary to find a sufficiently compact
notation. We illustrate it by considering first the simple case of a
product of two $\nabla$'s. This has two terms, one where the derivative in
the first $\nabla$ acts on the $\beta$'s in the second $\nabla$, and one
where both derivatives are free to act on whatever is on their right-hand side:
\bea
\nabla_l \nabla_k &=& \left( \sum_{n} \vec{\beta}_l^{(n)}
  \cdot \vec{\partial}^{(n)} \right) \left( \sum_{m} \vec{\beta}_k^{(m)}
  \cdot \vec{\partial}^{(m)} \right) \nn 
&=& \sum_{n,m}\left[
\beta_{li}^{(n)}  {\partial \beta_{kj}^{(m)} \over \partial c_i^{(n)}}
  {\partial \over \partial c_j^{(m)}}
+ \beta_{li}^{(n)} \beta_{kj}^{(m)} {\partial \over \partial c_i^{(n)}}
 {\partial \over \partial c_j^{(m)}}
 \right] \co 
\eea
where we have used the summation convention for repeated indices.
We denote these two terms by
\be
\label{eq:2prod}
[lk] = (l,k)+(l)(k) \fs
\ee
To each bracket corresponds one free derivative, and the brackets commute
by definition. Consider now a product of three $\nabla$'s, and construct it
by multiplying from the left the product of two $\nabla$'s with another
$\nabla$. The latter can either act on the other two $\nabla$'s or remain
free. In the notation just introduced this can be rephrased as follows: if
we add a new index from the left in the left-hand side of Eq.~(\ref{eq:2prod})
this can enter from the left on each of the brackets on the right-hand side of
Eq.~(\ref{eq:2prod}), or stay alone in its own bracket:
\be
\label{eq:3prod}
[jlk] = ([jl],k)+(k)(j,l)+(l)(j,k)+(j)(l,k)+(j)(l)(k) \fs
\ee
Using Eq.~(\ref{eq:2prod}) the first term could still be rewritten as
\be
([jl],k)=((j,l),k)+((j)(l),k) \fs
\ee
In general a bracket can have at most two arguments: the first one (if
present) is a $\nabla$, or a product thereof, whose free derivatives are all
acting on the $\nabla$ identified by the second argument. The derivative
corresponding to the latter $\nabla$ remains free.

Consider now a generic product of $k$ $\nabla$'s. It is easy to convince
oneself that one can write it as
\be
[l_1 \ldots l_k]=\sum_{
\stackrel{\mbox{\tiny all}}{\mbox{\tiny splittings}}}
  \prod_{\stackrel{\mbox{\tiny all}}{\mbox{\tiny
    subsets}}}([l_{i_1}\ldots l_{i_{j-1}}],l_j) \co
\label{eq:gprod}
\ee
and that a recursive use of the latter will generate all the terms in the
product. Note that the sum has to run over all possible splittings of the
set of $k$ numbers into subsets, and that the order of the $l_i$'s inside
the subsets has to be the same as in the original set.

Until now all the indices appearing in this notation referred to the
$l$-order, but it is useful also to introduce an index related to the
$v$-order. A given $\nabla_l$ contains derivatives with respect to coupling
constants of any $v$-order larger than $l$:
\be
\label{eq:nabla}
\nabla_l=\sum_{n=l}^\infty \vec{\beta}_l^{(n)}\cdot \vec{\partial}^{(n)}
\fs 
\ee
A $\nabla$ (or a product thereof), however, always acts on objects (the
$\cA^{(n)k}_{l}$) which are polynomials in the coupling constants
$c_i^{(n)}$, and which have a maximum $v$-order, such that only a finite
number of terms in the infinite sum (\ref{eq:nabla}) will play a role. The
$v$-order of a monomial in the $c_i^{(n)}$'s is defined as
\be
N_v\left[ \prod_i \left(c_{j_i}^{(n_i)} \right)^{k_i}  \right] = \sum_i n_i
k_i \fs  
\ee
Analogously, a product of derivatives will reduce the $v$-order of the
object it acts on by the amount 
\be
\Delta_v\left[ \prod_i \left(\partial_{j_i}^{(n_i)} \right)^{k_i}  \right]
= \sum_i n_i k_i \fs
\label{eq:Nvpartial}
\ee
A product of derivatives acting on a monomial gives zero if the $v$-order
of the latter is lower than the $\Delta_v$ of the derivatives. For a
polynomial in the $c_i^{(n)}$'s it is important to identify its maximum
$v$-order: for a $\cA^{(n)}_{lk}$ this is equal to $n-l$ (which is also its
minimum $v$-order). The first RGE relates $\cB^{(n)}_l$ to
$\cA^{(n)}_{l1}$: the $v$-order of $\cB^{(n)}_l$ is therefore also equal to
$n-l$. 

Eq.~(\ref{eq:nabla}) shows that the minimal $\Delta_v$ of $\nabla_l$ is
$l$. If we consider products of $\nabla$'s, their minimal $\Delta_v$ is
\be
\min \left( \Delta_v[l_1 \ldots l_k] \right) = \sum_i l_i \fs
\label{eq:Nvln}
\ee
It is convenient to write the brackets as sums of terms with a definite
$v$-order: 
\be
([l_1 \ldots l_k],j) = \sum_{n=l_1+\ldots+l_k+j}^\infty ([l_1 \ldots
l_k],j)_n \co
\ee
where we have made explicit the fact that the minimal $\Delta_v$ of a
bracket is equal to the sum of all the indices inside the bracket
(\ref{eq:Nvln}). 
Moreover, according to (\ref{eq:Nvpartial}), the $\Delta_v$ of a product of
brackets is equal to the sum of their $\Delta_v$'s:
\be
\Delta_v \Big[([\ldots],l_1)_{n_1} ([\ldots],l_2)_{n_2} \ldots
  ([\ldots],l_j)_{n_j} \Big] = \sum_{i=1}^j n_i \fs
\ee

\subsection{Highest pole equation}
The equation for the highest pole reads:
\be
n! \cA^{(n)n}_{n}=\nabla_1^{n-1} \beta_{1}^{(n)} \fs
\label{eq:hpe1}
\ee
Note that this is actually a set of equations for each of the components of
$\cA^{(n)n}_{n}$ and $\beta_{1}^{(n)}$:
\be
\cA^{(n)}_{nn}= \vec{a}^{(n)}_{nn} \cdot \vec{\cO}^{(n)} \; , \; \; \;
\cB_{1}^{(n)}=\vec{\beta}^{(n)}_1 \cdot \vec{\cO}^{(n)} \co
\ee
and that the operators $\cO^{(n)}_i$ just play the role of a basis of vectors,
and only allow us to write the equation more compactly.
In fact the content of the equations remains exactly the same if we
substitute $\cO^{(n)}_i \to \partial^{(n)}_i$. If we do
that, Eq.~(\ref{eq:hpe1}) gets rewritten as
\be
n! d_n = ([\underbrace{11 \ldots 1}_{n-1}];1)_n =
([1^{n-1}],1)_n \co 
\label{eq:hpe1a}
\ee
where we have directly used the notation with the brackets, and where we
have introduced the symbol $[1^{n-1}]$ for the product of $n-1$ $\nabla_1$.

In order to express fully explicitly the highest poles $\cA^{(n)}_{nn}$, or
equivalently the $d_n$, we now have to use Eq.~(\ref{eq:gprod}), and split
the $n-1$ $\nabla_1$ in all possible subsets:
\be
[1^{n-1}]_{n-1} = \sum_{\{n_i m_i=n-1\}} (n-1)! \prod_i
c_i \Big[ ([1^{n_i-1}],1)_{n_i} \Big]^{m_i} 
\co 
\label{eq:hpe2}
\ee
where $c_i$ is a combinatorial factor that we will discuss below.
Note that we have used explicitly the fact that the total $\Delta_v$ has to
be equal to $n$: this implies that all brackets can contribute only with
their minimal $\Delta_v$ -- according to Eq.~(\ref{eq:hpe1a}) their
contribution is equal to $n_i! d_{n_i}$.

Finally we have to discuss the factors $c_i$, which count how many times a
term is generated in the expansion of the product of $n-1$ $\nabla_1$'s. We
do this in the following steps:
\begin{enumerate}
\item 
We first permute the $n$ factors $\nabla_{1}$ in all possible ways, and get
a factor $(n-1)!$ (already written explicitly in Eq.~(\ref{eq:hpe2})). 
We can now just count the different splittings of $n-1$ into smaller
integers, and not consider the ordering. This however generates an
overcounting, which is compensated in the next two steps.
\item 
The ordering of the $\nabla_{1}$'s in each subset has to be like the original
ordering. To compensate for this overcounting we must include a factor $ 1 / 
(n_{i}!)^{m_{i}}$ in $c_i$.  
\item
The $m_{i}$ copies of the same subset are not distinguishable: $c_i$ must
also contain a factor $ 1/ m_{i}!$.
\end{enumerate} 
In total we get:
\be
c_i= {1 \over (n_i!)^{m_i} m_i!} \co
\ee
and finally
\be
n! d_n = \left( S_{n-1} ,1 \right)_n \co
\label{eq:hpe3}
\ee
where
\be
S_n := [1^n]_n= n! \sum_{\{n_i m_i=n\}} \prod_i {1 \over m_i!} d_{n_i}^{m_i}
\label{eq:Sn}
\ee
which is the result we were after.

\setcounter{equation}{0}
\section{Beyond the highest pole equation}
\label{sec:bhpe}
We will now consider the divergences $\cA^{(n)}_{lk}$ with $k < n$. The
starting point is the RGE (\ref{eq:k2l}) which we rewrite here for
convenience
\[
l \cA_{lk}^{(n)} =  \sum_{l'=1}^{l-k+1} \nabla_{l'} \cA_{l-l' \, k-1}^{(n)}
\qquad l=k, \ldots n, \; \; k = 2,\ldots,n \fs  
\]
The equation relates $\cA_{lk}^{(n)}$ to $\cA_{l'\,k-1}^{(n)}$, but if we
apply it recursively we end up relating it to $\cA_{l'\,1}^{(n)}=1/l'
\cB_{l'}^{(n)}$, in the following manner
\be
 \label{eq:mf1}
 \mA_{lk}^{(n)} = \sum_{\underline{l}} c_{\underline{l}} 
 \nabla_{l_{1}}\cdot...\cdot\nabla_{l_{k-1}} \cB_{l_{k}}^{(n)} \; ; \quad
  c_{\underline{l}} = \left( \prod_{j=1}^{k} \sum_{m=j}^{k} l_{m}
  \right)^{-1} 
  \co 
\ee
where the sum runs over all possible ordered $k$-ple  
$ \underline{l} =(l_{1},...,l_{k})$ with the property $ \sum_{i=1}^{k}
l_{i} = l$.
Using the notation introduced in the previous section we can rewrite this
as
\be
d^{(n)}_{lk}= \sum_{\underline{l}} c_{\underline{l}}([l_1....l_{k-1}],l_k)_n
\co
\label{eq:mf2}
\ee
where
\be
\cA^{(n)}_{lk}= \vec{a}^{(n)}_{lk} \cdot \vec{\cO}^{(n)} \; \Rightarrow \; \; \; 
d^{(n)}_{lk}= \vec{a}^{(n)}_{lk} \cdot \vec{\partial}^{(n)} 
\fs
\ee
If we want to make this equation more explicit we have to expand the bracket
$[l_1....l_{k-1}]$ according to Eq.~(\ref{eq:gprod}), and relate the
various brackets to $d$'s with lower $\hbar$-order. The scheme is
recursive, and allows one to go to as high an \ho as one wants, but having
a fully explicit formula to all orders like the one we had for the highest
pole looks very difficult. In order to illustrate what kind of difficulties
one faces beyond the leading poles we will now discuss the case of the
subleading poles.

In the case of the subleading poles ($k=n-1$) we have to deal with two
different $l$-order. Eq.~(\ref{eq:mf2}) can be written down as follows for
this case 
\bea
(n-1)! d^{(n)}_{n-1\, n-1} &=& ([1^{n-2}],1)_n \nn
n! d^{(n)}_{n\, n-1} &=& ([1^{n-2}],2)_n + \sum_{j=0}^{n-3} (n-1-j)([1^j 2
1^{n-3-j}],1)_n \; ,
\label{eq:subl1}
\eea
where
\be
[1^j 2 1^k]:= [\underbrace{1\ldots 1}_{j} 2 \underbrace{1 \ldots 1}_{k}]
\fs 
\ee
We first consider the first equation in (\ref{eq:subl1}): the right-hand
side has $\Delta_v=n$ and therefore can be split into two terms:
\be
([1^{n-2}],1)_n=([1^{n-2}]_{n-2},1)_n+([1^{n-2}]_{n-1},1)_n \fs
\ee
The first term can be written down explicitly: $[1^{n-2}]_{n-2}$ is just
$S_{n-2}$, whose expression is given in Eq.~(\ref{eq:Sn}). As for the
second term, it has the same structure as $S_{n-2}$, but one order of
$\Delta_v$ higher. It is useful to introduce a new symbol for $[1^n]_{n+1}$
\be
S_n^1 := [1^n]_{n+1} = n! \sum_{n_0=1}^n \left[ { 1 \over
  (n-n_0)!} d^{(n_{0}+1)}_{n_0 \, n_0} S_{n-n_0} \right] \co
\label{eq:Sn1}
\ee
where the latter expression can be obtained with the following reasoning:
one starts by expanding $[n]$ according to Eq.~(\ref{eq:gprod}), and
obtains 
\be
[1^n] = \sum_{\{n_i m_i=n \}} n! \prod_i {1 \over (n_i!)^{m_i} m_i!} \Big(
([1^{n_i-1}],1) \Big)^{m_i}
\fs
\label{eq:[n]}
\ee
The product $[1^n]$ contains terms of arbitrary $\Delta_v$, starting from
$n$, but here we are interested only in the part with $\Delta_v=n+1$. The
part with minimal $\Delta_v=n$ is obtained when all brackets have their
minimal $\Delta_v=n_i$ -- the part with $\Delta_v=n+1$ is obtained when
only one of the brackets has $\Delta_v=n_i+1$ and all others the minimal:
\be
[1^n]_{n+1} = \sum_{\{n_0+n_i m_i=n \}} {n! \over n_0!} ([1^{n_0-1}],1)_{n_0+1}
\prod_i {1 \over (n_i!)^{m_i} m_i!} \Big( ([1^{n_i-1}],1)_{n_i} \Big)^{m_i}
\fs
\label{eq:nn1}
\ee
After substituting all the brackets with the corresponding $d$'s, and
grouping together all the $d_n$'s into $S_i$'s one obtains the result in
Eq.~(\ref{eq:Sn1}). The expression for $d^{(n)n-1}_{n-1}$ can now be given
fully explicitly:
\be
(n-1)! d^{(n)}_{n-1 \, n-1}=(S_{n-2},1)_n+(S^1_{n-2},1)_n \fs
\ee
Note that this is again a recursive formula: $d^{(n)}_{n-1 \, n-1}$ is
expressed in terms of $d_m$'s with $m \leq n-2$ and $d^{(m)}_{m-1 \, m-1}$
with $m \leq n-1$.

We now come to the second equation in (\ref{eq:subl1}): the new object
which we have to deal with is $[1^j21^k]_{j+2+k}$. One can express this as
follows:
\bea
[1^j21^k]_{j+2+k}&=& \left. \sum_{j_1=0}^j  { j \choose j_1} \right[
  (S_{j_1},2)_{j_1+2} S_{j+k-j_1}  \nn
&+& \left. \sum_{k_1=1}^k  {k \choose k_1}
  ([1^{j_1}21^{k_1-1}],1)_{j_1+k_1+2} S_{j+k-j_1-k_1}   \right] \fs
\label{eq:121}
\eea
The derivation of this formula follows from the observation
that when we split the product on the left-hand side into subsets according
to Eq.~(\ref{eq:gprod}), only one of the subsets will contain a 2 -- the
various terms differ by the number of 1's to the left and right of the 2 in
the same subset. Moreover, since we are interested here only in the part
with minimal $\Delta_v$, what multiplies the subset with a 2 can be
expressed as a $S_n$. The combinatorial factors are then easily obtained.

We can now insert Eq.~(\ref{eq:121}) back into (\ref{eq:subl1}) and get
\bea
n! d^{(n)}_{n \, n-1}\!\!&\!\!=\!\!&\! \!(S_{n-2},2)_n +
\sum_{j=0}^{n-3}(n\!-\!1\!-\!j) \sum_{j_1=0}^j {j \choose j_1} 
((S_{j_1},2)S_{n-3-j_1},1)_n  \\
&\!\!+\!\!&\!\!\!\sum_{j=0}^{n-4}\!(n\!-\!1\!-\!j)\! \! \sum_{j_1=0}^j
 \!\!{j \choose j_1} \!\! \sum_{k_1=1}^{n\!-\!3\!-\!j}\!\! {n\!-\!3\!-\!j
   \choose k_1} \! (([1^{j_1}2 1^{k_1-1}],1) S_{n-3-j_1-k_1},1)_n \fs
\nonumber
\eea
One can again use Eq.~(\ref{eq:121}) to simplify further the last term and
gets
\bea
n! d^{(n)}_{n \, n-1}\!\!&\!\!=\!\!&\! \!(S_{n-2},2)_n +
\sum_{j=0}^{n-3}(n\!-\!1\!-\!j) \sum_{j_1=0}^j {j \choose j_1} 
((S_{j_1},2)S_{n-3-j_1},1)_n  \\
&\!\!+\!\!&\!\!\!\sum_{j=0}^{n-4}\!(n\!-\!1\!-\!j)\! \! \sum_{j_1=0}^j
 \!\!{j \choose j_1} \!\! \sum_{k_1=1}^{n\!-\!3\!-\!j}\!\! {n\!-\!3\!-\!j
   \choose k_1}\!\! \sum_{j_2=0}^{j_1}\!\! {j_1 \choose j_2} \times \nn
&& \qquad \qquad \qquad \times
(((S_{j_2},2)S_{j_1+k_1-1-j_2},1)S_{n-3-j_1-k_1},1)_n \nn  
&\!\!+\!\!&\!\!\!\sum_{j=0}^{n-5}\!(n\!-\!1\!-\!j)\! \! \sum_{j_1=0}^j
 \!\!{j \choose j_1} \!\! \sum_{k_1=1}^{n\!-\!3\!-\!j}\!\! {n\!-\!3\!-\!j
   \choose k_1}\!\! \sum_{j_2=0}^{j_1}\!\! {j_1 \choose j_2}\!\!
\sum_{k_2=1}^{k_1-1} \! {k_1\!-\!1 \choose k_2} \times \nn
&& \quad \times
((([1^{j_2}2 1 ^{k_2-1}],1)S_{j_1+k_1-1-j_2-k_2},1) S_{n-3-j_1-k_1},1)_n
\fs 
\eea
Using Eq.~(\ref{eq:121}) as many times as it is needed to fully eliminate
the terms containing $[1^j21^k]$. In this manner one obtains an expression
for the subleading poles which is fully explicit, again 
in the sense of a recursive formula: to have leading and subleading poles
at \ho $n$ one must have already worked out all leading and subleading
poles of \ho $n' < n$, and insert these as vertices in one- and two-loop
diagrams. 

\setcounter{equation}{0}
\section{Renormalization of the effective action}
\label{ap:gammas}

In Section~\ref{sec:eom} we have shown that in order to have the sum of 1PR
contributions to the generating functional automatically finite, order by
order in the loop expansion, one should systematically renormalize the
effective actions $\Gamma_n[\phi,J]$ also away from the classical solution. 
This can always be done because the divergences of $\Gamma_n$ must be
local: to renormalize it is enough to include in the counterterm basis also
terms that vanish at the EoM. In this section we discuss in some more
details how this can be done, and consider the case of $\Gamma_n$, assuming
that all $\Gamma_k$ with $k<n$ have already been renormalized also away
from the EoM. This implies that $Z_n^\mathrm{1PR}$ is finite, and therefore
that $\bar \Gamma_n$ is also finite. We can write a $\Gamma_n$ which is
finite at the EoM as 
\be
\Gamma_n= \sum_{i=1}^n \eps^{-i} \Gamma_{n\; i} +\Gamma_n^f(\eps) \qquad
\bar \Gamma_{n \; i}=0 \co
\ee
with $\Gamma_n^f(0)$ finite. Our argument applies to all $\Gamma_{n \; i}$
and, to simplify the notation we drop the subscript. Since $\Gamma$
vanishes at the EoM, we can write it as 
\be
\Gamma=\sum_n \hat c_n X^n_r S_0^r =: X_r S_0^r  \co
\label{eq:gamma}
\ee
with $S_0^r=0$ the classical EoM.
We are now interested to study its behaviour away from the EoM, and can
conveniently do this with a Taylor expansion:
\be
\phi=\phi_{cl}+\xi \Rightarrow \Gamma=\bar \Gamma^a \xi_a+ O(\xi^2) \fs
\ee
We want to ensure that $\Gamma$ vanishes also away from the EoM, and
therefore that
\be
\bar \Gamma^a= \bar X_r \Delta^{ra} = 0 \co
\ee
which can only be true if $\bar X_r = 0$. This condition can be easily
satisfied by properly adjusting the coefficients $\hat c_n$ in front of the
counterterms that vanish at the EoM (\ref{eq:gamma}). $X_r$,
however, may still be different from zero away from the EoM:
\be
X_r = X_{rs} S_0^s \co 
\ee
which implies
\be
\Gamma={1 \over 2} \bar \Gamma^{ab} \xi_a \xi_b + O(\xi^3) \co \qquad 
\bar \Gamma^{ab}=\bar X_{rs} \Delta^{ra} \Delta^{sb} \fs
\ee
$\Gamma^{ab}=0$ implies $\bar X_{rs}=0$, which can be obtained by tuning
the coefficients of the counterterms that vanish quadratically at the EoM,
and in turn means
\be
X_{rs} = X_{rst} S_0^t \co
\ee
and so on. Note that the expansion in $\xi$ of a term that vanishes at
the EoM contains powers of the inverse propagator $\Delta$: when inserted
in 1PR graphs such vertices will generate local, possibly divergent
contributions. By changing the coefficients in front of the EoM terms I can
shift local contributions from $\Gamma_n$ to $Z_n^\mathrm{1PR}$.

We stress that the procedure for the renormalization of $\Gamma_n$ outlined
above does not need to go on forever. First of all because at every finite
\ho only a finite number of conditions have to be imposed,
cf. Eq.~(\ref{eq:eom_cond}). Moreover, in any QFT, the EoM must reduce, in
a well defined limit, to the free-field ones 
\be
S_0^i=(\Box+M^2) \phi^i+\sigma^{ij} \phi_j = 0 \fs
\ee
For dimensional reasons the powers of $S_0^i$ which can be contained in a
counterterm of \ho $n$ is bounded. For example in a renormalizable theory
not more than one power of $S_0^i$ can appear in a counterterm of any \ho.
In a nonrenormalizable theory higher-dimensional interactions are
suppressed by powers of an energy scale. This ordering is usually reflected
in the $\hbar$-ordering, such that at each order in $\hbar$ only a limited
power of $S_0^i$ can appear. In CHPT, e.g., the EoM are of chiral order
two, such that at \ho one (chiral order four) not more than two powers of
$S_0^i$ are allowed: the condition
\be
\Gamma_1^i=\Gamma_1^{ij}=0 \co
\ee
ensures that $\Gamma_1$ is finite, even away from the classical solution.
In general, for $\Gamma_n$ the chiral counting implies that there are $n+1$
conditions to be imposed:
\be
\Gamma_n^{i_1}=\Gamma_n^{i_1 i_2} =\Gamma_n^{i_1 \ldots i_{n+1}}=0 \fs
\ee

\end{document}